\documentclass[journal]{IEEEtai}

\usepackage[colorlinks,urlcolor=blue,linkcolor=blue,citecolor=blue]{hyperref}

\usepackage{color,array}

\usepackage{graphicx}

\usepackage{microtype}
\usepackage{subfigure}
\usepackage{booktabs} % for professional tables
\usepackage{amsmath,amssymb}
\DeclareMathOperator{\E}{\mathbb{E}}

\usepackage{hyperref}

\usepackage{times}
\usepackage{soul}
\usepackage{url}
\usepackage[utf8]{inputenc}
\usepackage{amsthm}
\usepackage{booktabs}
\usepackage{algorithm}
\usepackage{algorithmic}
\usepackage{todonotes}
\usepackage[hang,flushmargin]{footmisc} \usepackage{xspace}

\hypersetup{
    colorlinks=true,
    citecolor=blue
}

\newif\ifdraft
\draftfalse
\newcommand{\saurabh}[1]{\ifdraft{\todo[inline,color=cyan!40]{saurabh: #1}}\fi}

\newcommand{\somali}[1]{\ifdraft{\todo[inline,color=red!40]{somali: #1}}\fi}
\newcommand{\pranjal}[1]{\ifdraft{\todo[inline,color=olive!40]{pranjal: #1}}\fi}
\definecolor{coquelicot}{rgb}{1.0, 0.22, 0.0}

\newcommand{\ie}{{\em i.e.,}\xspace}

\newcommand{\remove}[1]{}

\begin{document}

\title{Federated Action Recognition on Heterogeneous Embedded Devices}

\author{Pranjal Jain*, Shreyas Goenka*, Saurabh Bagchi, Biplab Banerjee, Somali Chaterji
% \thanks{This paragraph of the first footnote will contain the date on which you submitted your paper for review. It will also contain support information, including sponsor and financial support acknowledgment. For example, ``This work was supported in part by the U.S. Department of Commerce under Grant BS123456.'' }
\thanks{\textbf{*Pranjal Jain and Shreyas Goenka contributed equally to this paper.}}
\thanks{Pranjal Jain is with the Electrical Engineering Department, Indian Institute of Technology Bombay, Mumbai, India (email: pranjal0000@gmail.com).}
\thanks{Shreyas Goenka is with the Electrical Engineering Department, Indian Institute of Technology Bombay, Mumbai, India (email: shreyasgoenka7@gmail.com).}
\thanks{Saurabh Bagchi is with the Electrical and Computer Engineering and Computer Science Departments, Purdue University, West Lafayette, USA (email: sbagchi@purdue.edu).}
\thanks{Biplab Banerjee is with the Centre of Studies in Resources Engineering (CSRE), Indian Institute of Technology Bombay, Mumbai, India (email: bbanerjee@iitb.ac.in).}
\thanks{Somali Chaterji is with the Agricultural and Biological Engineering Department, Purdue University, West Lafayette, USA (email: schaterji@purdue.edu).}
% \thanks{This paragraph will include the Associate Editor who handled your paper.}
}

\markboth{Journal of IEEE Transactions on Artificial Intelligence}
{Jain \MakeLowercase{\textit{et al.}}: IEEE Journal of IEEE Transactions on Artificial Intelligence}

\maketitle

\begin{abstract}
Federated learning allows a large number of devices to jointly learn a model without sharing data. In this work, we enable clients with limited computing power to perform action recognition, a computationally heavy task. We first perform model compression at the central server through knowledge distillation on a large dataset. This allows the model to learn complex features and serves as an initialization for model fine-tuning. The fine-tuning is required because the limited data present in smaller datasets is not adequate for action recognition models to learn complex spatio-temporal features. Because the clients present are often heterogeneous in their computing resources, we use an asynchronous federated optimization and we further show a convergence bound. We compare our approach to two baseline approaches: fine-tuning at the central server (no clients) and fine-tuning using (heterogeneous) clients using synchronous federated averaging. We empirically show on a testbed of heterogeneous embedded devices that we can perform action recognition with comparable accuracy to the two baselines above, while our asynchronous learning strategy reduces the training time by 40\% relative to synchronous learning.
\end{abstract}

\begin{IEEEImpStatement}
In order to enable edge devices to perform action recognition using limited computing power, we have developed a federated learning framework that also takes into account the heterogeneity of resources present on each embedded device. CCTV cameras, which are critical to security applications, often require the classification of abnormal activity. Since privacy is crucial, the video data cannot always be transferred to a central server. Hence, the training needs to occur at the edge devices themselves. In order to ensure that downtime on certain devices does not affect the rest of the system, in this paper, we use asynchronous updates. The framework in this paper can be used in a variety of applications, including, but not limited to, industry, defense and government applications. 
\end{IEEEImpStatement}


\section{Introduction}
\label{sec:introduction}

\saurabh{1. What is the problem? \\
2. How has prior work tried to solve it? \\
3. What is the shortcoming of that work? \\
4. What is the main idea of our solution? \\
5. What are the key features that our solution achieves? \\
6. How do we evaluate and demonstrate that our solution works? \\
7. What are the 3 claims to novelty that we make in this paper? \\
Add an overview figure with blocks showing how our solution works. 
}
\pranjal{Done. Please do a pass over this}
Action recognition has been a long-standing and actively pursued problem in the computer vision community due to its practical applications in areas such as surveillance, semantic video retrieval and multimedia mining. Video action recognition involves the identification of actions from video clips. This may seem like an extension of image classification to multiple frames: predicting the action in each frame and aggregating the predictions. However, the success of image classification on datasets such as ImageNet~\cite{deng2009imagenet} has not been replicated in video classification. Several challenges exist in video recognition such as huge computational costs, capturing long spatio-temporal contexts, and designing classification architectures. Human action recognition approaches can be categorized into visual sensor-based, non-visual sensor-based, and multi-modal categories~\cite{yurur2014survey,ranasinghe2016review}. The main difference between visual and other categories is the form of the sensed data. The visual data are captured in the form of 2D/3D images or video whilst others capture the data in the form of a 1D signal~\cite{ranasinghe2016review}.

Apart from fully-supervised methods~\cite{girdhar2017actionvlad, carreira2017quo, diba2018temporal, girdhar2019video}, several papers discuss few-shot~\cite{zhang2020few,kumar2019protogan} mechanisms and zero-shot learning mechanisms~\cite{brattoli2020rethinking,mandal2019out}. As the name suggests, in few-shot learning, the number of training examples is limited, and in zero-shot learning, we use test samples from classes that were not observed during training. 
\somali{Put a one-liner for zero-shot as well.}
The datasets available for the task have variable sizes; some such as the Kinetics~\cite{kay2017kinetics} have sufficient data to train large models, whereas others like HMDB51~\cite{kuehne2011hmdb} and UCF101~\cite{soomro2012ucf101} result in model overfitting due to limited data. 
\somali{put a reference for this deficit of these latter two datasets; I am guessing this is the one .. if so, we are good: ~\cite{hara2017learning}}
We also empirically see here that ResNet
%\somali{which ResNet architecture is this}
training resulted in significant overfitting for UCF-101 and HMDB-51 but not for Kinetics~\cite{hara2017learning}: ResNet-18 trained from scratch on HMDB51 achieved a per-clip accuracy of 17.1\% whereas a model pretrained on the Kinetics dataset and finetuned on HMDB51 achieved a per-clip accuracy of 56.4\%~\cite{hara2018can}.
\somali{even with 56.5\%, the accuracy looks low to a reader who is not familiar with these datasets so add a 1-2 line explanation}
As datasets grow in size and complexity, an additional limiting factor is that of constraints such as computation, time required, and costs associated with training~\cite{li2019budgeted}. These constraints become crucial in edge devices, which typically have limited computation and storage and unpredictable network connectivity to a central server~\cite{bonawitz2019towards}. Limited device storage in turn implies that clients can accommodate only small local datasets, and as seen in previous work~\cite{hara2018can}, this results in inaccurate models.

Although computer vision algorithms have achieved high accuracy on several tasks, it comes at the price of large computational costs~\cite{feng2019computer}. Federated learning aims to leverage the computing power and data of a multitude of clients, often with significant heterogeneity in terms of computational power and network bandwidth, to build models. Federated learning is also motivated by the desire for privacy preservation~\cite{bonawitz2017practical}, eliminating the risk of sensitive data transfer to a central server~\cite{blanchard2017machine,xu2019verifynet,mugunthan2020privacyfl}. 
%Federated learning eliminates this risk and aims to provide protection against malicious attacks~\cite{blanchard2017machine,xu2019verifynet,mugunthan2020privacyfl}.
%\somali{Can you expand on privacy preservation in a couple lines.}
%\pranjal{Done}
Furthermore, the technique is most often implemented using the synchronous approach, which assumes that clients proceed in lockstep. Particularly in our target domain of embedded devices, a synchronous approach can get bottlenecked due to a few straggler clients as has been argued before~\cite{chai2019towards,bonawitz2019towards,chai2020tifl}.
%\saurabh{Give citation to prior work, e.g., see related-work.tex and if we have results from our own experiments.}
%\pranjal{done}
This stems from the heterogeneity among the clients in resources and datasets, and variation in network connectivity~\cite{xie2019asynchronous}. 
% The heterogeneity of data results in non-IID datasets on the clients and synchronous algorithms such as FedAvg~\cite{mcmahan2017communication} slow down convergence because of stragglers~\cite{li2019convergence}. 
\somali{I don't think you can claim that heterogeneity in data results in non-IID, I like the line of thought where you talk about the differential bias in non-IIDness and the need for asynchronous algorithms ... but the mapping from heterogeneity to non-IID does not pass muster}
Hence, in this work we choose % SB (1/31/21): Use present tense. So "we choose to use" not "we chose to use"
to use an asynchronous federated learning optimization. 
% Further, we choose to use models that have been initialized using knowledge distillation on the central server and on large datasets to fine-tune on embedded devices and on small datasets.

\begin{figure}[ht]
\vskip 0.2in
\begin{center}
\centerline{\includegraphics[width=\columnwidth]{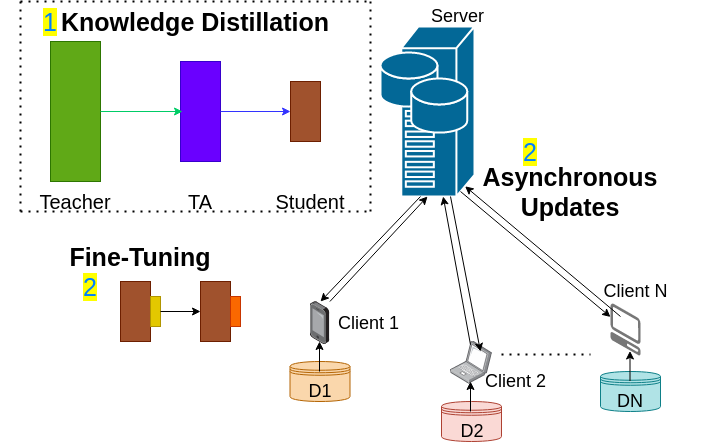}}
\caption{The central server first performs knowledge distillation using the Kinetics dataset: from teacher to teaching assistant (TA) and from TA to student. The students (compressed models) are then fine-tuned on the smaller dataset using an asynchronous federated optimization.}
% \saurabh{Take a look at overview\_handdrawn.pdf. Purple colored text is instruction, not for putting in the figure.}
% \pranjal{Done}
% Resolved
\label{fig:overview}
\end{center}
\vskip -0.2in
\end{figure}

% SB (2/5/21): Following text is there in related work
% Previous work include review of various distributed computer vision algorithms~\cite{radke2010survey} and computer vision as a cloud service~\cite{agrawal2015cloudcv}.
% Previous work include an adaptive control algorithm that determines the best tradeoff between local updates and global aggregation under a given resource constraint~\cite{wang2019adaptive}, model training in a network of heterogeneous edge devices taking into account communication costs~\cite{lim2020federated} and a method for straggler acceleration by dynamically masking neurons~\cite{xu2019elfish}. Aso-fed~\cite{chen2019asynchronous} presents an online learning algorithm that updates the central model in an asynchronous manner that tackles challenges associated with both varying computational loads at heterogeneous edge devices and stragglers.
% FedVision~\cite{liu2020fedvision} is the first paper to discuss setting up federated learning for object detection from images. However, they use the synchronous FedAvg~\cite{mcmahan2017communication} algorithm in their work, which is adversely affected by stragglers and heterogeneity. 
% Another paper discusses a learning rate schedule~\cite{li2019budgeted} based on available resources, but it does not consider that datasets may be distributed amongst edge devices.
\saurabh{This prior work discussion is not relevant as it is very basic about federated learning. What is relevant is federated learning on edge devices and asynchronous federated learning. See for example the file related-work.tex.}
\pranjal{done}
In this paper we discuss enabling action recognition in edge devices through federated learning. An overview of our approach is given in Figure~\ref{fig:overview}. Considering that edge devices have limited computing and memory, the datasets present on them are small. To overcome this, we first distill knowledge from large models, pretrained on the larger Kinetics dataset, at the central server. We use two approaches to this knowledge distillation: one in which the teacher directly teaches the student~\cite{hinton2015distilling}, and the second approach, where there is an intermediate teaching assistant (TA)~\cite{mirzadeh2020improved}. We observe that the latter approach is superior, albeit it comes at the cost of higher training time, and we adopt this approach in our pipeline. In the original work on TA, the setting was image classification. Here, we show through empirical evaluation that introducing a TA between the student and the teacher improves the accuracy for action recognition through videos. We further investigate the the effects of using additional TAs in our pipeline. While  the  increase in per-clip, top-1 accuracy is appreciable when one TA is introduced, using additional TAs does not produce any considerable improvement in accuracy. On the other hand, the training time increases sharply as more TAs are added.
\saurabh{The original paper already tells us this. Did we have to do anything novel to apply this? Did we get any insights from this approach?}
\pranjal{done}

The advantage of knowledge distillation is twofold: first, it compresses the model---advantageous for limited computing, and second, it initializes a model for fine-tuning. Since we use heterogeneous embedded devices as clients, we incorporate an {\em asynchronous} federated optimization at the fine-tuning stage. We also show that under certain assumptions, the algorithm has a convergence bound. To the best of our knowledge, no previous work tackles the problem of performing action recognition from video clips, via federated learning. In addition to this, we compare our approach to two baselines: \textit{first}, fine-tuning on the central server (no clients), and \textit{second}, fine-tuning on clients using synchronous federated averaging. We use as evaluation metrics accuracy and time. For the accuracy, we use top-1 accuracy at the clip level\footnote{A clip consists of 8 video frames.}, and at the entire video level. The finer granularity (clip-level) allows us to model (near) real-time action recognition and is a more challenging metric. For the metric of time, we measure both training and inference times. 

In summary, our paper makes the following contributions: \\
% \begin{enumerate}
\noindent 1. Our work is the first to enable action recognition, a computationally heavy task, on edge devices, which have limited computing and memory resources and hence can only accommodate small-sized datasets on them.

\noindent 2. In order to circumvent the problem of limited data, we apply knowledge distillation, with and without an intermediate teaching assistant (TA) using the Kinetics dataset, and observe that the latter is superior. Furthermore, we experiment by adding more TAs but observe that the increase in accuracy is negligible. Thus, our work expands on the original work to use TAs~\cite{mirzadeh2020improved} to the significantly heavier task of activity recognition.
    % \somali{Isn't this second contribution a given?}
    % \pranjal{done}
    % Knowledge distillation has two benefits: model compression and allowing students to learn complex spatiotemporal features from larger models 

\noindent 3. The edge devices used as clients are heterogeneous, mimicking many real-world settings. We show a convergence bound for asynchronous training. We implement and run experiments on a heterogeneous testbed of embedded devices and show that the asynchronous training has significant advantages over synchronous, while the accuracy achieved is comparable to the centralized baseline. 
% \somali{Another contribution could be related to the latency guarantees that are backed by our mathematical proofs.}
% \pranjal{Don't have any proof for latency guarantees. But I have changed contribution 2 so we have 3 claims to novelty}
% \end{enumerate}

\section{Related Work}
\label{sec:related}
FedVision~\cite{liu2020fedvision} is the first paper to discuss setting up federated learning for object detection from images. Current approaches, which build object detection models on centrally located datasets, suffer from the high cost of transmitting video data and privacy concerns that are alleviated using federated learning. They provide a platform to support the development of object detection models. However, FedVision uses the synchronous FedAvg~\cite{mcmahan2017communication} algorithm in their work, which is adversely affected by stragglers and heterogeneity. Another approach~\cite{yu2019federated}, which tackles object detection, uses Kullback-Leibler divergence (KLD) to measure the divergence of the weight distribution in training and proposes FedAvg with Abnormal Weights Clip to relief the influence of non-IID data, enabling
higher performance with non-IID data. Federated learning has also been used to solve problems in medical imaging~\cite{kaissis2020secure}. To the best of our knowledge, no work so far has tackled the problem of action recognition using federated learning.

Federated distillation~\cite{jeong2018communication} follows an online version of knowledge distillation, known as co-distillation (CD)~\cite{anil2018large}. In CD, each device treats itself as a student, and sees the mean
model output of all the other devices as its teacher’s output. Furthermore, to rectify non-IID data of on-device ML can be corrected by obtaining the missing local data samples at each device from the other devices. This can induce significant overhead, so FAug is proposed which generates the missing data on each device. They empirically found that their approach yields lower overhead and better accuracy for image classification on MNIST~\cite{lecun1998gradient}. Human action recognition approaches can be categorized into visual sensor-based, non-visual sensor-based and multi-modal categories~\cite{yurur2014survey,ranasinghe2016review}. So far, federated learning has only been incorporated into federated learning using wearable sensors~\cite{sozinov2018human,ek2020evaluation}.

For federated learning on heterogeneous devices, previous work include an adaptive control algorithm that determines the best trade-off between local updates and global aggregation under a given resource constraint~\cite{wang2019adaptive}, model training in a network of heterogeneous edge devices, taking into account communication costs~\cite{lim2020federated}, and a method for straggler acceleration by dynamically masking neurons~\cite{xu2019elfish}. Aso-fed~\cite{chen2019asynchronous} presents an online learning algorithm that updates the central model in an asynchronous manner, tackling challenges associated with both varying computational loads at heterogeneous edge devices and stragglers. There has been a sparse but rapidly growing work in federated learning at edge devices, driven by the increasing numbers of such edge devices~\cite{nikouei2018smart,lin2018edgespeechnets,wang2019adaptive}. Training and deploying smaller yet accurate networks is of interest in edge devices. Knowledge distillation compresses the knowledge of a large and computationally expensive model (often an ensemble of neural networks) to a single computationally efficient neural network.
% The idea of knowledge distillation is to train the small model, the student, on a transfer set with soft targets provided by the large model, the teacher.
The fact that these edge devices are often constrained in terms of local resources (compute, memory, and storage) as well as network resources (low bandwidth connections, intermittent connectivity) has given rise to fruitful areas of inquiry in communication-efficient federated learning~\cite{reisizadeh2020fedpaq, sattler2019robust, mills2019communication}, asynchronous learning to deal with stragglers~\cite{smith2017federated, xu2019elfish}, approximate models and computation~\cite{zhang2018shufflenet, wu2019fbnet, han2020ghostnet}, and knowledge distillation to create more succinct models~\cite{jang2020knowledge, matsubara2020head}. 

% Knowledge distillation for action recognition; using a TA
% In the last few years, there has been a push towards deploying artificial intelligence at the edge~\cite{nikouei2018smart,lin2018edgespeechnets}. This is to address requirements such as user privacy, limited network connectivity and computation resources. Federated learning is a well-recognized collaborative learning technique which leverages individual device data and computation resources such as limited memory~\cite{smith2017federated,bonawitz2019towards}.
% tackles challenges  associated with both varying computational loads at heterogeneous edge devices and stragglers.\\

When knowledge distillation is done via intermediate TAs, the accuracy for image classification is seen to increase~\cite{mirzadeh2020improved}. Other applications of knowledge distillation in computer vision include those for semantic segmentation~\cite{he2019knowledge,liu2019semantic}, image classification~\cite{xie2020self}, and action recognition~\cite{thoker2019cross}.
% SB (2/4/21): [CTM (Comment to myself)]: thoker2019cross distills knowledge from one teacher network to a small ensemble of student networks. Their claim to fame is that the training set has one modality and the test set has a differion~\cite{huang2018multimodal}. \\
To date, the applications of federated learning to computer vision have been limited. FedVision~\cite{liu2020fedvision} provides a tool for the development of object detection frameworks. 
% The drawbacks in this work is the usage of the synchronous FedAvg algorithm~\cite{mcmahan2017communication}, which is affected by heterogeneous devices and stragglers.
% Apart from this, the use of federated learning has been explored in the medical imaging domain~\cite{kaissis2020secure,li2019privacy}.
{\em Distinct from all prior work, we are the first to perform federated learning for action recognition, a computationally heavier task than those previously attempted, leveraging heterogeneous edge devices.}
% To the best of our knowledge, no previous work discusses federated learning used for action recognition from videos. 

\section{Background and Problem Statement}
\label{sec:background}
\subsection{Architecture} 
\begin{figure}[ht]
% \vskip 0.2in
\begin{center}
\scalebox{0.25}{\centerline{\includegraphics[width=\columnwidth]{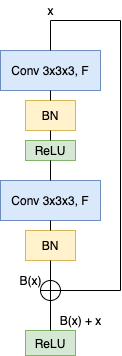}}}
\caption{ResNet building block}
\label{fig:block}
\end{center}
\vskip -0.2in
\end{figure}

ResNets~\cite{he2016deep,hara2017learning} introduced shortcut connections that bypass a signal from one layer to the next. The connections pass through the gradient flows of networks from later layers to early layers, and ease the training of very deep networks. The ResNets that we use here performs 3D convolution and 3D pooling. Figure~\ref{fig:block} shows the basic building block that we use while building our models. Here $F$ refers to the number of convolution filters. The dimensions of $B(x)$ and $x$ may be different, therefore, we use 1$\times$1$\times$1 convolutions to match dimensions. In our experiments, we use ResNet-18, ResNet-26, and ResNet-34 which are derived from the building block. For the knowledge distillation experiments, the teacher we use is ResNet-34, the student is ResNet-18 and the teaching assistant is ResNet-26. Subsequent fine-tuning is performed on the distilled ResNet-18 model.

% \begin{figure}[ht]
% % \vskip 0.2in
% \begin{center}
% \scalebox{1.0}{\centerline{\includegraphics[width=\columnwidth]{figures/arch.png}}}
% \caption{ResNet architectures derived from the building block}
% \label{fig:arch}
% \end{center}
% \vskip -0.2in
% \end{figure}

\subsection{Knowledge Distillation}

Knowledge distillation leverages information present in a larger ``teacher" model to train a ``smaller" student model. The soft probabilities (or logits) output from the teacher convey additional information to the student~\cite{hinton2015distilling}: rather than just using training labels, students can learn more by understanding the relation between different classes. The additional information from the soft probabilities can help the student network identify the decision boundaries better if multiple classes have high soft probabilities.\\
Concretely, given an input $\mathbf{x}$, the teacher model computes a vector of scores (or logits)  $\pmb{z}^t(\mathbf{x})$ $= [z_1^t(\mathbf{x}), z_2^t(\mathbf{x}) \ldots z_K^t(\mathbf{x})]$ for $K$ classes. We define the knowledge distillation loss $L_{KD}$ as the Mean Squared Error between the logits from the teacher model $\pmb{z}^t$ and the student model $\pmb{z}^s$, \ie $L_{KD} = \left \| \pmb{z}^t(x) - \pmb{z}^s(x) \right \|^2$. The overall loss function is a combination of two loss functions,  $L = \alpha L_{cls} + (1-\alpha) L_{KD}$ where $L_{cls}$ is the conventional cross-entropy loss which is computed for the predictions made by the student and the ground truth corresponding to the input $\mathbf{x}$.

The teacher model cannot effectively transfer its knowledge to the student if the size gap between them is large~\cite{mirzadeh2020improved}. To alleviate this, the knowledge distillation is done in steps and an intermediate-sized model, the teaching assistant (TA), is introduced. Thus, in the first round of a \textit{teacher-TA-student} knowledge distillation algorithm, the TA distils knowledge from the teacher model, with the loss function $L_1 = \alpha L_{cls}^{TA} + (1-\alpha) L_{KD}^{TA}$. Subsequently, the student distils the knowledge that is gained by the TA and is trained with the loss function $L_2 = \alpha L_{cls}^{S} + (1-\alpha) L_{KD}^{S}$. Here $\pmb{z}^{TA}$ refers to the vector of logits computed by the TA, $L_{KD}^{TA}= \left \| \pmb{z}^t(x) - \pmb{z}^{ta}(x) \right \|^2$, $L_{KD}^{S} = \left \| \pmb{z}^{ta}(x) - \pmb{z}^{s}(x) \right \|^2$, and $L_{cls}^{TA}$ and $L_{cls}^{S}$ are the cross-entropy losses for the TA and student model respectively, and these are calculated considering the ground truth to be the output of the teacher for the input $\mathbf{x}$.
\subsection{Fine Tuning}
Transfer learning is used to improve a learner from one domain by transferring information from a related domain. One possible need for transfer learning occurs when there is a limited supply of target training data~\cite{wang2018deep}. Action recognition models require large datasets to learn complex spatiotemporal features~\cite{hara2018can}. In this work, we aim to enable edge devices with limited memory and computation abilities to perform action recognition. For example, amongst the edge devices we are using in this experiments, the NVIDIA Jetson AGX Xavier, the most well-resourced device type, has a 32GB storage and cannot store large datasets like Kinetics-400 which requires an approximate storage space of 400GB.

Next, we formally define transfer learning. A domain $\Psi$ is defined by a feature space $\chi$ and a marginal probability distribution $P(X)$, where $X=\{\mathbf{x}_1,\mathbf{x_2},...,\mathbf{x_n}\} \in\chi$. For a given domain $\Psi$, a task $\Gamma$ is defined by a label space $Y$ and a predictive function $f(\cdot)$. The training data consists of the pairs $\{\mathbf{x}_i,y_i\}$ where $\mathbf{x}_i\in \chi$ and $y_i \in Y$. Hence, we define a domain $\Psi=\{\chi,P(X)\}$ and a task $\Gamma=\{Y,f(\cdot)\}$. We define the source task $\Gamma_S$, target task $\Gamma_T$, source predictive function $f_{S}(\cdot)$, target predictive function $f_{T}(\cdot)$, source domain $\Psi_S$ and target domain $\Psi_T$. Formally, transfer learning is the process of improving the predictive function $f_T(\cdot)$ using the related information from $\Psi_S$ and $T_S$, where $\Psi_S \neq \Psi_T$ or $\Gamma_S \neq \Gamma_T$.  

\saurabh{Present (1) Background - activity recognition DNNs, knowledge distillation, fine tuning, federated learning}
\subsection{Problem Formulation}
We consider a federated learning setup with $n$ devices. Our aim is to find the parameters $w$ that solve $\min_{w} F(w)$. Here, $F(w) = \frac{1}{n}\sum_{k=1}^{n} \E [l(w;d^k)]$ where $d^k$ is data sampled from local data $D^k$ on the $k$-th device, and $l(\cdot;\cdot)$ is a user specified loss function.\\
% \[ \min_{w} F(w)\]
The training takes $E$ global epochs. In the $t^{th}$ epoch, the central server receives an updated $w_{new}$ from a client. 
% SB (2/1/21): I think there should be a subscript for the client. Thus: receives $W_{new}^i$ from the $i$-th client. 
The model parameters are then averaged $w_t = (1-\beta_{t})w_{t-1}+\beta_{t} w_{new}$, where $\beta \in (0,1)$ is the mixing hyperparameter. 
% SB (2/1/21): I believe there is some inconsistency in this notation. LHS is a global variable and the RHS aggregation should be done across all clients. The \beta values should be client specific. 
Because of the presence of stragglers, we define {\em ``staleness''} as $t-\tau$ where $\tau$ is the epoch on the local device, and adaptively calculate the $\beta_{t}$. On the $i^{th}$ device, after receiving the global model $w_t$, the local optimization is $\min_{w} \E[l(w;d^k) + \frac{\theta}{2} \|w-w_t\|^2] = \min_{w}\E[g_{w_{t}}(w;d^k)]$. Here $\theta$ is the regularization hyperparameter. 
% SB (2/1/21): Explain the above equation. 
All norms used in the paper are $L_2$-norms. The model trains for a number of local iterations, where that number $\in [H_{min}, H_{max}]$ with a learning rate $\eta$. We define the imbalance ratio $\lambda = \frac{H_{max}}{H_{min}}$. $w_{t, h}^i$ refers to the weights from epoch $t$ on device $i$ after $h$ training iterations.
% SB (2/1/21): I do not understand "The model trains ..."
The number of iterations is specified by the server to the clients and it can do this based on local knowledge (such as, the desire to avoid hotspot congestion at the server) or based on client knowledge (such as, the compute resources at each edge device).
% \saurabh{The problem formulation right now only captures the asynchronous training. It also needs to formally specify the knowledge distillation and then the fine tuning. The background on all these concepts must come before the problem formulation.}
\section{Algorithmic and System Details}
\label{sec:design}

\saurabh{Here provide the details of our math formulation and solution, algorithm, and our system. A good analysis is to show bound on the accuracy relative to the simple baseline of centralized training using the same dataset.}
\subsection{Asynchronous Federated Optimization}
The global server and the clients conduct training in an asynchronous manner; the client starts training and sends updates to the server as soon as it is done. Because the clients are heterogeneous, there is no synchronization between the various devices. 
\begin{algorithm}[tb]
  \caption{Asynchronous Federated Learning}
  \label{alg:async}
\begin{algorithmic}
  \STATE {\bf Server}
%   \REPEAT
  \STATE Initialize $w_0$.
  \FOR{$t = 1$ {\bfseries to} $E$}
  \STATE Receive $(w_{new},\tau)$ from any client
  \STATE $\beta_t \gets \beta \times s(t-\tau)$,$s(\cdot)$ is a function of the staleness
  \STATE $w_t \gets (1-\beta_{t})w_{t-1}+\beta_{t} w_{new}$
%   \IF{$x_i > x_{i+1}$}
%   \STATE Swap $x_i$ and $x_{i+1}$
%   \STATE $noChange = false$
%   \ENDIF
  \ENDFOR
%   \UNTIL{$noChange$ is $true$}
    \STATE {\bf Client}
    \FOR{$k \in \{1,\ldots,n\}$ in parallel}
    \STATE receive global model and time stamp $(w_t,t)$
    \STATE $\tau \gets t $ , $w_{\tau,0}^{k} \gets w_{t}$
    \STATE Define $g_{w_{t}}(w;d)=l(x;z)+\frac{\theta}{2}\|w-w_t\|^2$
    \FOR{local iteration $h \in \{1,\ldots,H_{\tau}^k\}$}
    \STATE Sample $d_{\tau,h}^k$ from $D^k$
    \STATE $w_{\tau,h}^k \gets w_{\tau,h-1}^k-\eta \nabla g_{w_{t}}(w_{\tau,h-1}^k;d_{\tau,h}^k)$
    % SB (2/1/21): Shouldn't \tau be updated here?
    \ENDFOR
    \STATE Send $(w_{\tau,H_{\tau}^k}^k,\tau)$ to the server
    \ENDFOR
\end{algorithmic}
\end{algorithm}
The $k^{th}$ client performs training with a learning rate $\eta$ using data $d_{\tau,h}^k$, which is randomly sampled from its local dataset $D^{k}$. $H_{\tau}^k$ is the number of local iterations performed. $E$ is the total number of global epochs.\\
% The server is a multi-threaded system that connects to the clients through sockets. All the threads run independently and in parallel to each other. 
% SB (2/1/21): This kind of low-level detail is not needed here, if at all in the paper this can come in the experimental evaluation. 
Because of delays by clients in sending the updates due to various reasons like low battery, high latency, or low bandwidth, different clients have different values of staleness, $t-\tau$. Intuitively, a large staleness means that the global model is more accurate because it has been trained more. 
% SB (2/1/21): I do not understand above sentence. A high staleness means that this particular client is sending an update which is stale. This should mean that if the server is reliant on this client for an accurate update to the model, then it will suffer. 
Hence, aggregating with stragglers results in errors being introduced. To mitigate this, we use a function $s(t-\tau)$, which adaptively changes the mixing parameter $\beta_t = \beta \times s(t-\tau)$ --- the weight allocated to $w_{new}$ in the aggregation. The general form of this function is that $s(t-\tau)=1$ when $t=\tau$ and it monotonically decreases with increase in $(t-\tau)$.

\onecolumn
\subsection{Convergence Analysis}
We make the following assumptions in our analysis. Assumptions 1 and 2 are standard; typical examples are $l_2$ regularized linear regression, logistic regression, and softmax classification.\\
\textbf{Assumption 1} Assume that $F(\cdot)$ is L-smooth; $\forall v,w, F(v) \leq F(w) + \langle \nabla F(w),v-w \rangle + \frac{L}{2} \|v-w\|^2$.\\
\textbf{Assumption 2} Assume that $F(\cdot)$ is $\mu$-weakly convex; there exists a function $G(\cdot)$ which is convex, such that $\forall v,G(v)=F(v)+\frac{\mu}{2}\|v\|^2$.\\
\textbf{Assumption 3} The staleness of stragglers $t-\tau$, where $t$ represents the current global epoch and $\tau$ represents the local epoch, is bounded $t-\tau \leq K$.\\
\textbf{Assumption 4} Assume that for a device $k$, the square norm of the gradients are bounded. $\| \nabla l(w;d^k)\|^2 \leq B_{1}^2$ and $\| \nabla g_{w_{t}}(w;d^k)\|^2 \leq B_{2}^2$.\\
\textbf{Theorem} Under assumptions 1 through 4, for any constant $\epsilon>0$, and choosing $\theta$ large enough such that $\theta>\mu$ and $-(1+2 \theta + \epsilon)B_2^2 + (\theta^2 - \frac{\theta}{2}) \|w_{\tau, h-1} - w_{\tau}\|^2 \geq 0$  $\forall w_{\tau, h}, w_{\tau}$  ,  and learning rate $\eta < \frac{1}{L}$. After $E$ global updates, algorithm~\ref{alg:async} converges to a critical point,
\[ \min_{t=0}^{E-1} E\|\nabla F(w_t)\|^2 \leq \frac{\E[F(w_{0}) - F(w_{E})]}{\beta \eta \epsilon E H_{min}}+\mathcal{O}(\frac{\eta \lambda^3 H_{min}^2}{\epsilon})\]\[+\mathcal{O}(\frac{\beta K \lambda}{\epsilon})+\mathcal{O}(\frac{\eta K^2 \lambda^2 H_{min}}{\epsilon})\]\[+\mathcal{O}({\frac{\beta^2 \eta K^2 \lambda^2 H_{min}}{\epsilon}})\]
Here $F(w_t)$, the objective function, represents the training loss in epoch $t$. Intuitively, when we upper bound the square of $L2$-norm of the gradient, $\nabla F(\cdot)$ approaches zero, and since we use gradient steps to update weights, the algorithm converges. 

Choosing the learning rate $\eta = \frac{1}{\sqrt{E}}$, and having the number of training epochs (E) approach infinity, the asymptotic upper bound becomes:
\[\lim_{E\rightarrow \infty} \min_{t=0}^{E-1} E\|\nabla F(w_t)\|^2 \leq \mathcal{O}(\frac{\beta K \lambda}{\epsilon})\]
The above asymptotic upper bound can be made arbitrarily close to zero by increasing $\epsilon$.

\textbf{Brief Outline of Proof} 

Take $\theta$ large enough such that $-(1+2 \theta + \epsilon)B_2^2 + (\theta^2 - \frac{\theta}{2}) \|w_{\tau, h-1} - w_{\tau}\|^2 \geq 0$ \\
We write $\nabla g_{w_\tau}(w_{\tau, h-1}^i, d_{\tau, h}^i)$ as $\nabla g_{w_\tau}(w_{\tau,h-1})$ for simplicity.\\
$\langle \nabla G_{w_\tau}(w_{\tau, h-1}), \nabla g_{w_\tau}(w_{\tau, h-1}) \rangle - \epsilon \| \nabla F(w_{\tau, h-1}) \|^2$\\
$=\langle \nabla F(w_{\tau, h-1})+\theta(w_{\tau,h-1}-w_{\tau}), \nabla l(w_{\tau, h-1})+\theta(w_{\tau,h-1}-w_{\tau}) \rangle - \epsilon \| \nabla F(w_{\tau, h-1}) \|^2$\\
$=\langle \nabla F(w_{\tau, h-1}), \nabla l(w_{\tau, h-1}) \rangle + \theta \langle \nabla F(w_{\tau, h-1})+\nabla l(w_{\tau, h-1}),w_{\tau, h-1} - w_{\tau} \rangle + \theta^2\|w_{\tau, h-1} - w_{\tau}\|^2- \epsilon \| \nabla F(w_{\tau, h-1}) \|^2$\\
$\geq -0.5\| \nabla F(w_{\tau, h-1}) \|^2 - 0.5\|\nabla l(w_{\tau, h-1})\|^2-0.5\theta \| \nabla F(w_{\tau, h-1})+\nabla l(w_{\tau, h-1})\|^2 - 0.5\theta \|w_{\tau, h-1} - w_{\tau}\|^2 + \theta^2\|w_{\tau, h-1} - w_{\tau}\|^2 - \epsilon \| \nabla F(w_{\tau, h-1}) \|^2$\\
$\geq -0.5\| \nabla F(w_{\tau, h-1}) \|^2 - 0.5\|\nabla l(w_{\tau, h-1})\|^2 - \theta \| \nabla F(w_{\tau, h-1}) \|^2 -\theta \nabla l(w_{\tau, h-1})\|^2 - 0.5\theta \|w_{\tau, h-1} - w_{\tau}\|^2 + \theta^2\|w_{\tau, h-1} - w_{\tau}\|^2 - \epsilon \| \nabla F(w_{\tau, h-1}) \|^2$
$\geq -(1+2 \theta + \epsilon)B_2^2 + (\theta^2 - \frac{\theta}{2}) \|w_{\tau, h-1} - w_{\tau}\|^2 \geq 0$\\
\begin{equation}
\label{1}
\implies \eta \langle \nabla G_{w_\tau}(w_{\tau, h-1}), \nabla g_{w_\tau}(w_{\tau, h-1}) \rangle \leq \eta\epsilon\| \nabla F(w_{\tau, h-1}) \|^2
\end{equation}
Using $\tau-(t-1) \leq K$, we have $\|w_\tau - w_{t-1}\|^2 \leq \|(w_{\tau}-w_{\tau-1}) + \ldots + (w_{t-1}-w_{t-1})\|^2$\\
$\leq K\|(w_{\tau}-w_{\tau-1})\|^2 + \ldots + K\|(w_{t-1}-w_{t-1})\|^2 \leq \beta^2\eta^2K^2\lambda^2H_{min}^2\mathcal{O}(B_2^2)$
\begin{equation}
\label{2}
    \|w_\tau - w_{t-1}\|^2 \leq \beta^2\eta^2K^2\lambda^2H_{min}^2\mathcal{O}(B_2^2)
\end{equation}
Also, $\|w_\tau - w_{t-1}\| \leq \|(w_{\tau}-w_{\tau-1}) + \ldots + (w_{t-1}-w_{t-1})\| \leq \|w_{\tau}-w_{\tau-1}\|+\ldots+\|w_{t-1}-w_{t-1}\| \leq \beta\eta K\lambda H_{min}\mathcal{O(B_2)}$
\begin{equation}
\label{3}
    \|w_\tau - w_{t-1}\| \leq \beta\eta K\lambda H_{min}\mathcal{O}(B_{2}) 
\end{equation}
Consider L-smoothness and convexity assumptions,\\
$\mathbb{E}[F(w_{\tau,h})-F(w_*)] \leq \mathbb{E}[G_{w_{\tau}}(w_{\tau,h})-F(w_*)]$\\
$\leq G_{w_{\tau}}(w_{\tau,h})-F(w_*) - \eta \mathbb{E}[\langle \nabla G_{w_\tau}(w_{\tau, h-1}), \nabla g_{w_\tau}(w_{\tau, h-1}) \rangle] + 0.5L\eta^2\mathbb{E}[\|\nabla g_{w_\tau}(w_{\tau, h-1})\|^2]$\\
$\leq F(w_{\tau,h})-F(w_*)+0.5\theta \|w_{\tau, h-1} - w_{\tau}\|^2- \eta \mathbb{E}[\langle \nabla G_{w_\tau}(w_{\tau, h-1}), \nabla g_{w_\tau}(w_{\tau, h-1}) \rangle] + 0.5L\eta^2\mathbb{E}[\|\nabla g_{w_\tau}(w_{\tau, h-1})\|^2]$\\
$\leq F(w_{\tau,h})-F(w_*)- \eta \mathbb{E}[\langle \nabla G_{w_\tau}(w_{\tau, h-1}), \nabla g_{w_\tau}(w_{\tau, h-1}) \rangle]+0.5L\eta^2B_{2}^2 + \frac{\theta}{2}\eta^2\lambda^2 H_{min}^2 B_{2}^2$\\
$\leq F(w_{\tau,h})-F(w_*)-\eta \mathbb{E}[\langle \nabla G_{w_\tau}(w_{\tau, h-1}), \nabla g_{w_\tau}(w_{\tau, h-1}) \rangle] + \eta^2\mathcal{O}(\theta \lambda^2 H_{min}^2 B_{2}^2)$\\
$\leq F(w_{\tau,h})-F(w_*) - \eta\epsilon\| \nabla F(w_{\tau, h-1}) \|^2+\eta^2\mathcal{O}(\theta \lambda^2 H_{min}^2 B_{2}^2)$ (Using~\ref{1})\\
Rearrange and telescope,\\
\begin{equation}
\label{4}
    \mathbb{E}[F(w_{\tau,h})-F(w_{\tau})] \leq -\eta \epsilon \sum_{h=0}^{H-1}\mathbb{E}\|\nabla F(w_{\tau,h})\|^{2} + \eta^2\mathcal{O}(\theta\lambda^3H_{min}^3B_{2}^2)
\end{equation}
Using L-smoothness,\\
$F(w_{\tau})-F(w_{t-1}) \leq \langle \nabla F(w_{t-1}),w_{\tau}-w_{t-1}\rangle + \frac{L}{2}\|w_{\tau}-w_{t-1}\|^2$\\
$\leq \|\nabla F(w_{t-1})\| \|w_{\tau}-w_{t-1}\| + \frac{L}{2}\|w_{\tau}-w_{t-1}\|^2$\\
$\leq B_1\beta\eta K \lambda H_{min} \mathcal{O}(B_2) + \frac{L}{2}\beta^2\eta^2K^2\lambda^2H_{min}^2\mathcal{O}(B_2^2)$(Using~\ref{2} and~\ref{3})\\
$\leq  B_1\beta\eta K \lambda H_{min} \mathcal{O}(B_2) +\beta^2\eta^2K^2\lambda^2H_{min}^2\mathcal{O}(B_2^2)$\\
\begin{equation}
    \label{5}
    F(w_{\tau})-F(w_{t-1}) \leq \beta\eta K \lambda H_{min} \mathcal{O}(B_1B_2) +\beta^2\eta^2K^2\lambda^2H_{min}^2\mathcal{O}(B_2^2)
\end{equation}
Consider,\\
$\mathbb{E}[F(w_{t})-F(w_{t-1})] \leq \mathbb{E}[G_{w_{t-1}}(w_{t})-F(w_{t-1})]$\\
$\leq \mathbb{E}[(1-\beta)G_{w_{t-1}}(w_{t-1})+\beta G_{w_{t-1}}(w_{\tau,h})-F(w_{t-1})]$\\
$\leq \mathbb{E}[\beta(F(w_{\tau,h})-F(w_{t-1}))+\frac{\beta\theta}{2}\|w_{\tau, h} - w_{t-1}\|^2]$
$\leq \beta\mathbb{E}[F(w_{\tau,h})-F(w_{t-1}))] + \beta\theta\|w_{\tau, h} - w_{\tau}\|^2+\beta\theta\|w_{\tau} - w_{t-1}\|^2$\\
$\leq \beta\mathbb{E}[F(w_{\tau,h})-F(w_{t-1}))] + \beta\theta \beta^2\eta^2K^2\lambda^2H_{min}^2\mathcal{O}(B_2^2)$ (Using~\ref{2})\\
Using~\ref{4} and~\ref{5},\\
$\mathbb{E}[F(w_{t})-F(w_{t-1})]$\\
$\leq -\beta\eta\epsilon \sum_{h=0}^{H-1}\mathbb{E}\|\nabla F(w_{\tau,h})\|^{2} + \eta^2\mathcal{O}(\theta\lambda^3H_{min}^3B_{2}^2)+\beta^2\eta K \lambda H_{min} \mathcal{O}(B_1B_2) +\beta^3\eta^2K^2\lambda^2H_{min}^2\mathcal{O}(B_2^2)+\beta\eta^2K^2\lambda^2H_{min}^2\mathcal{O}(B_2^2)$\\
Note that $H_{t}^{'}$ is the number of local epochs applied in iteration $t$. Rearrange terms to get,\\
$\sum_{h=0}^{H_{t-1}^{'}}\mathbb{E}\|\nabla F(w_{\tau,h})\|^{2} \leq \frac{\mathbb{E}[F(w_{t})-F(w_{t-1})]}{\beta \eta \epsilon} + \frac{\eta\lambda^3 H_{min}^3 \mathcal{O}(B_2^2)}{\epsilon} + \frac{\beta K \lambda H_{min}\mathcal{O}(B_1B_2)}{\epsilon} + \frac{\beta^2 \eta K^2\lambda^2H_{min}^2\mathcal{O}(B_2^2)}{\epsilon}+\frac{\eta K^2 \lambda^2 H_{min}^2\mathcal{O}(B_2^2)}{\epsilon}$\\
By Telescoping and taking total expectation, after T global epochs, we have\\
$\min_{t=0}^{E-1} E\|\nabla F(w_t)\|^2 \leq \frac{1}{\sum_{t=1}^{\tau}H_{t}^{'}}\sum_{t=1}^{\tau}\sum_{h=0}^{H_{t-1}^{'}}\mathbb{E}\|\nabla F(w_{\tau,h})\|^{2}$\\
$\leq \frac{\E[F(w_{0}) - F(w_{E})]}{\beta \eta \epsilon E H_{min}}+\mathcal{O}(\frac{\eta \lambda^3 H_{min}^2}{\epsilon}) +\mathcal{O}(\frac{\beta K \lambda}{\epsilon})+\mathcal{O}(\frac{\eta K^2 \lambda^2 H_{min}}{\epsilon})+\mathcal{O}({\frac{\beta^2 \eta K^2 \lambda^2 H_{min}}{\epsilon}})$\\
This concludes the proof for the convergence bound of the asynchronous federated algorithm. In the experimental evaluation, we tune hyperparameters, such as $\beta$ in Figure~\ref{fig:b_HMDB51} and observe how the convergence varies.
\saurabh{Now explain the implications of this convergence rate. What are the most important parameters controlling convergence? How does this compare to prior results?}
\twocolumn
\section{Experimental Evaluation}
\label{sec:evaluation}

% \saurabh{Here for starters write down the set of experiments you will do and for each put a placeholder showing the table (put the column names) or a figure (show the X and Y axes and the expected curves). Put down what baselines we will compare to.}

% \somali{One table showing training times for: fine tuning at the server, knowledge distillation with and without TA, training federated learning.}

% \somali{A second table showing inferencing times for: different categories of clients.}

% \somali{A plot showing accuracy vs \# epochs, control parameter is degree of asynchronous nature.}

% \pranjal{All done}

The central server in the following experiments has an NVIDIA Tesla V100S 32GB GPU. We use a variety of clients to demonstrate that our asynchronous federated optimization is robust to heterogeneous edge devices: \textit{NVIDIA Jetson Nano}, which has 4GB memory, and a 128-core Maxwell GPU; \textit{NVIDIA Jetson TX2}, which has 8GB memory, and a 256-core Pascal GPU; \textit{NVIDIA Jetson Xavier NX}, which has a 8GB memory, and a 384-core Volta GPU with 48 Tensor cores; and NVIDIA Jetson AGX Xavier, which has 32GB memory and 512-core Volta GPU. The Kinetics~\cite{kay2017kinetics} dataset, which we use for knowledge distillation, is present at the central server, and we conduct experiments on two datasets for fine-tuning: HMDB51~\cite{kuehne2011hmdb} and UCF101~\cite{soomro2012ucf101}. This data is distributed amongst the clients. The Kinetics dataset contains 400 human action classes, with at least 400 video clips for each action. Each clip lasts for around 10s and is taken from a different YouTube video. The dataset has 306,245 videos, and is divided into three splits: one for training, having 250–1000 videos per class; one for validation, with 50 videos per class; and one for testing, with 100 videos per class. The HMDB51 dataset contains 51 classes and a total of 3,312 videos.The UCF101 dataset consists of 101 classes and over 13k clips (27 hours of video data).

In this paper, we use two metrics for evaluation of our pipeline: per-clip and per-video accuracy. Similar to a previous study~\cite{hara2018can}, we use the {\bf per-clip top-1} accuracy as an evaluation metric, \ie if we consider a 10 second video at 24 fps and 8 frames per clip are used, we get 30 clips. The top-1 refers to considering the model prediction to be accurate if the top class it outputs matches the label of the video. Most previous studies report the per-video accuracy by taking the mean of class scores output from all clips in a video and comparing it to the ground truth. We choose one of the evaluation metrics to be per-clip granularity level because it indicates how noisy the videos in a dataset are and it opens up scope for future work in which predictions can be made on clips rather than using the entire video --- the per-clip evaluation also allows for real-time predictions. We also use the {\bf per-video top-1} accuracy, in which we take the mean of class scores output from all clips in a video and compare it to the ground truth

\subsection{Knowledge Distillation}
\begin{figure}[h!]
% \vskip 0.2in
\begin{center}
\centerline{\includegraphics[width=0.9\columnwidth]{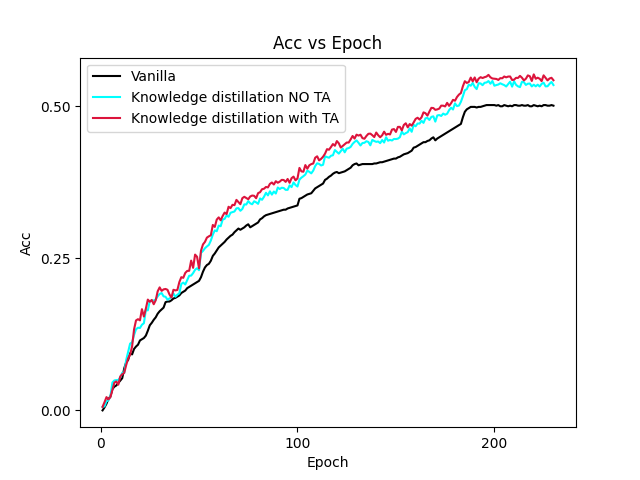}}
\caption{Top-1 accuracy on the Kinetics validation dataset for 3 experiments: 1. Training a ResNet-18 from scratch (Vanilla); 2. Knowledge distillation from ResNet-34 to ResNet-18 (Knowledge Distillation with no TA); 3. Knowledge distillation with ResNet-34 Teacher, ResNet-26 TA, and ResNet-18 Student (Knowledge Distillation with TA).}
\label{fig:kd}
\end{center}
\vskip -0.2in
\end{figure}

In the first stage of our pipeline, we perform knowledge distillation from a larger model, trained on the Kinetics dataset. We compare three approaches in order to validate using knowledge distillation with an intermediate TA. For these experiments, we use a batch size of 128, learning rate $\eta=0.1$, and an SGD optimizer with a weight decay 0.001 and momentum 0.9. In the first experiment, we train a ResNet-18 model from scratch on the Kinetics dataset and the per-clip top-1 accuracy achieved is 50.2\%. Using knowledge distillation, the accuracy is improved to 53.8\% when we distill directly from ResNet-34 to ResNet-18, and 54.6\% when we use a ResNet-26 as the intermediate TA between the teacher and student. From Figure~\ref{fig:kd}, it is evident that using a distilled ResNet-18 is better than using a ResNet-18 trained from scratch. There is a counter pull from the training time --- the KD approach (not counting the time to train the ResNet-34) takes 43\% longer than training from scratch (the ResNet-18). This can be explained by the fact that ``Train from scratch" includes only forward-backward passes on ResNet-18 with optimization using only cross-entropy loss. On the other hand, KD involves forward passes on the larger ResNet-34, forward-backward passes on ResNet-18, and optimization on ResNet-18 using a combination of both cross-entropy loss and the MSE on the logits (recall that we are fine-tuning only the last FC layer). This timing result is consistent with prior works that report on the timing performance of knowledge distillation~\cite{hinton2015distilling, sun2019patient}. 

We further investigate using multiple TAs. From Table~\ref{tab:train}, we see that the introduction of one TA increases the train time from 44 hours 58 minutes to 55 hours 23 minutes and the corresponding increase in per-clip accuracy is 0.8\%. Hence, there is a trade-off between increased training time and increased accuracy. Furthermore, the introduction of a TA almost always increases accuracy but the optimal number of TAs and size of each is an open research question~\cite{mirzadeh2020improved}. Additionally, TAs are used to bridge the gap between the student and teacher: by using a ResNet-26 between ResNet-34 and ResNet-18 we already accomplish this. If the gap between the teacher and student were larger, using additional TAs would be of benefit at the expense of increased computation and train time required. In order to reduce the train times and achieve comparable accuracy to the baselines, we use one TA.

\begin{figure}[ht]
% \vskip 0.2in
\begin{center}
\scalebox{1.0}{\centerline{\includegraphics[width=\columnwidth]{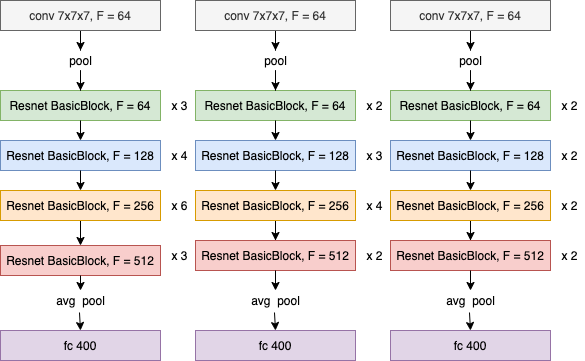}}}
\caption{ResNet-34, ResNet-26, and ResNet-18 architectures derived from the basic building block}
\label{fig:arch}
\end{center}
\vskip -0.2in
\end{figure}

We further investigate the effects of using additional TAs in our pipeline.
\begin{table}[h!]
\caption{Knowledge distillation is performed by varying the number of intermediate Teaching Assistants (TAs).}
\label{tab:kd}
% \vskip 0.15in
\begin{center}
\begin{small}
\begin{sc}
\resizebox{\columnwidth}{!}{
\begin{tabular}{lcccr}
\toprule
\# TAs & Epochs & Time (hrs, mins) (Increase) & Per-Clip Accuracy\\
\midrule
0    & 200 & 44 h 58 m (0\%) & 53.8\%\\
1 & 200 & 55 h 23 m (23.2\%) & 54.6\%\\
2  & 200 &  69 h 35 m (54.7\%) & 54.8\%\\
3 & 200 & 85 h 47 m (90.8\%) & 54.9\%\\
\bottomrule
\end{tabular}}
\end{sc}
\end{small}
\end{center}
\vskip -0.1in
\end{table}
In all these experiments, distillation is performed from teacher ResNet-34 to student ResNet-18. In the first experiment, we do not use any TAs. In the next experiment, we use ResNet-26 as a TA. The third experiment is performed using two TAs: ResNet-28 and ResNet-24, and in the final experiment, we use three TAs: ResNet-30, ResNet-26, and ResNet-22. The architectures have been outlined in Figure~\ref{fig:arch}. From Table~\ref{tab:kd}, we see that while the increase in \textbf{per-clip top-1} accuracy is appreciable when one TA is introduced, using additional TAs does not produce any considerable improvement in accuracy. The training time increases sharply as more TAs are added. Hence, in the subsequent stages in our pipeline, we chose to use a single TA. 

\begin{figure}[h!]
% \vskip 0.2in
\begin{center}
\centerline{\includegraphics[width=0.9\columnwidth]{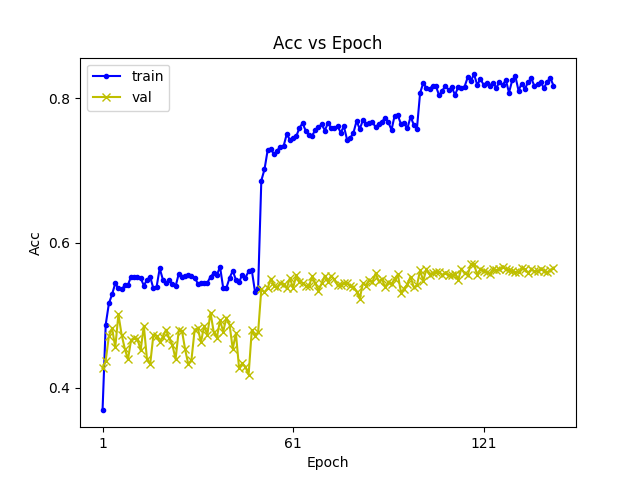}}
\caption{ResNet-18, distilled from ResNet-34, via a ResNet-26 Teaching assistant, and fine-tuned on HMDB51. The fine-tuning is performed at the central server without any clients.}
% \saurabh{Take a look at overview\_handdrawn.pdf. Purple colored text is instruction, not for putting in the figure.}
% \pranjal{Done}
% Resolved
\label{fig:central_HMDB51}
\end{center}
\vskip -0.2in
\end{figure}

\begin{figure}[h!]
% \vskip 0.2in
\begin{center}
\centerline{\includegraphics[width=0.9\columnwidth]{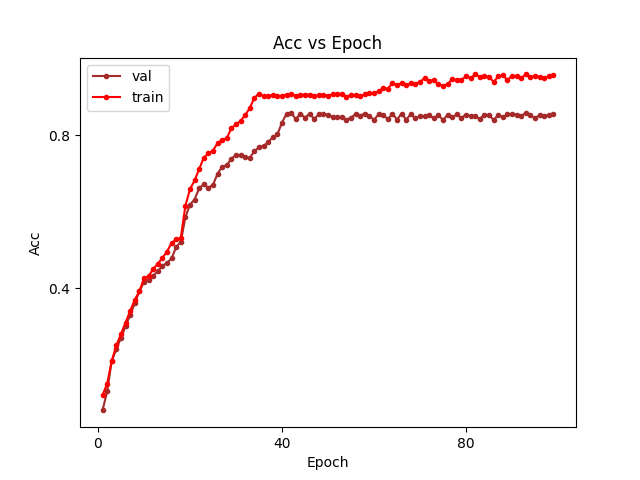}}
\caption{ResNet-18, distilled from ResNet-34, via a ResNet-26 Teaching assistant, and fine-tuned on UCF101, performed at the central server without any clients.}
% \saurabh{Take a look at overview\_handdrawn.pdf. Purple colored text is instruction, not for putting in the figure.}
% \pranjal{Done}
% Resolved
\label{fig:central_UCF101}
\end{center}
\vskip -0.2in
\end{figure}

For the rest of the experiments, we perform fine-tuning, by reinitializing the fully connected layer --- the last layer in the ResNet-18 model. The ResNet-18 being used is the model distilled from ResNet-34 (trained on the Kinetics dataset) via a ResNet-26 TA. The first baseline experiment that we perform is fine-tuning on the central server when no clients are present; the per-clip top-1 accuracy is shown in Figure~\ref{fig:central_HMDB51} for HMDB51 and Figure~\ref{fig:central_UCF101} for UCF101. A model trained in this manner achieves a per-clip top-1 accuracy of 57.3\% on HMDB51. In contrast, ResNet-18 trained from scratch on HMDB51 achieves a per-clip accuracy of only 17.1\%. This is consistent with several prior results that have indicated the difficulty of achieving high accuracy in activity recognition on the relatively small HMDB51 dataset~\cite{hara2018can, liu2017sparse}. 

\subsection{Transfer Learning}
The Kinetics-400 dataset requires an approximate disk space of 400GB to store. Amongst the edge devices we are using in these experiments, the most well-endowed, NVIDIA Jetson AGX Xavier has only 32GB storage. Hence, edge devices can only accommodate smaller-sized datasets on them\footnote{One may argue that adding cheap external storage such as through Flash cards can alleviate this problem. However, reading from external storage is orders of magnitude slower than reading from internal storage and will thus increase the training time to an infeasible level.}. In this section, we use the HMDB51 dataset and the UCF101 for evaluation. The HMDB51 which has a size of 2,062MB and is distributed amongst the clients in such a way that requires approximately 500MB of storage space on each client. The UCF101 is 6.9GB and each client has about 1.725GB of data. Once we have distilled knowledge from the larger ResNet-34, trained on the Kinetics dataset to the ResNet-18 architecture via a TA, the next step is to fine-tune on the smaller dataset; \ie HMDB51 or UCF101.

\begin{figure}[h!]
% \vskip 0.2in
\begin{center}
\centerline{\includegraphics[width=0.9\columnwidth]{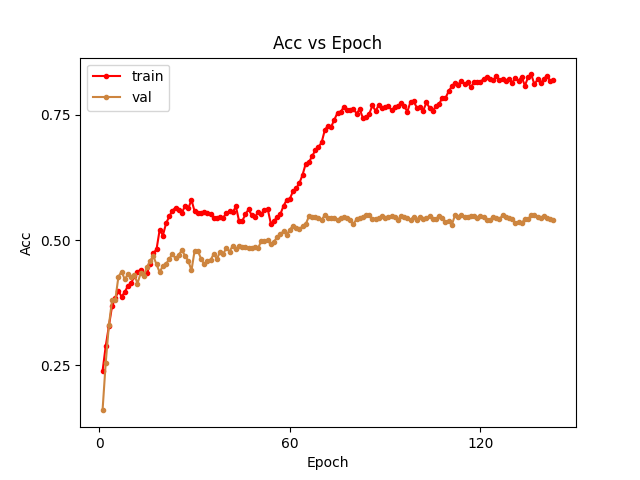}}
\caption{Fine-tuning on HMDB51 using a synchronous FedAvg algorithm. The clients connected to the central server are heterogeneous in their computational resources. For comparison, ResNet-18 trained from scratch on HMDB51 achieves a per-clip accuracy of 17.1\%.}
\label{fig:fedavg_HMDB51}
\end{center}
\vskip -0.2in
\end{figure}

\begin{figure}[h!]
% \vskip 0.2in
\begin{center}
\centerline{\includegraphics[width=0.9\columnwidth]{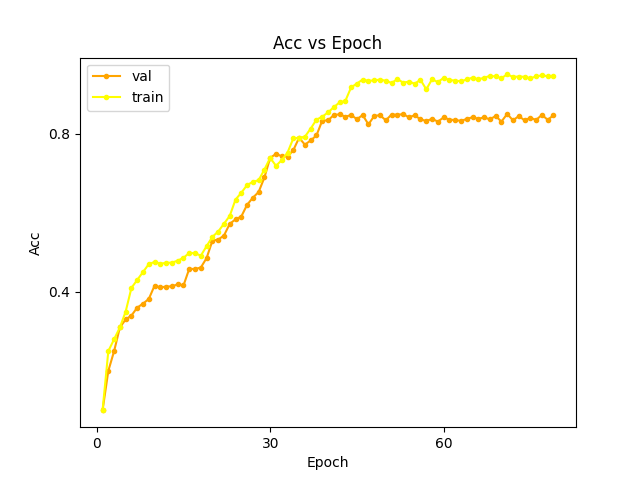}}
\caption{Fine-tuning on UCF101 using a synchronous FedAvg algorithm. The accuracy shown is the per-clip video accuracy.}
\label{fig:fedavg_UCF101}
\end{center}
\vskip -0.2in
\end{figure}

In the transfer learning experiments, we fine-tune only the last fully connected layer. We observed that the accuracy reduces as we increase the number layers that we fine-tune. In baseline experiments, we use a synchronous federated optimization, with 4 clients, one of each type (recall there are 4 kinds of embedded devices in our testbed). The accuracy \textit{vs}. number of global aggregations (epochs) is shown in Figure~\ref{fig:fedavg_HMDB51} for HMDB51, and Figure~\ref{fig:fedavg_UCF101} for UCF101. For each aggregation that the central server performs, the number of local epochs is 3. We use a batch size of 8, learning rate $\eta=0.001$, SGD optimizer with momentum 0.9. Each clip has 8 frames, which are taken from the corresponding video. 
% SB (2/5/21): Taken from the same video?
Although increasing the number of frames per clip would have increased the accuracy, doing so causes the clients to throw a CUDA out of memory error.

From Table~\ref{tab:train}, refer to the HMDB51 experiments. We see that the time required for a synchronous optimization is 10 hours and 54 minutes. In contrast, the asynchronous federated algorithm takes only 6 hours and 31 minutes, a 40\% decrease. This can be attributed to the clients having different computing resources, and hence requiring different amounts of time to complete the local epochs as given in Table~\ref{tab:traintime}. While the synchronous algorithm has to wait for the slowest client to send its update, the asynchronous algorithm continues its optimization. A similar effect is observed in the case of UCF101.
% SC062021: New text
One may wonder that it is beneficial to use the approach of fine tuning at the server without any clients (for HMDB51 and UCF101) and thus not having to use our approach. This alternate method runs into the problem that it does not leverage federated learning, which has its traditional benefits of scaling to a large number of clients (and thus not needing heavyweight server) and preserving privacy of client data. The same argument applies to why we would not want to train for the Kinetics data from scratch (this would obviously have to be done at the server).

\begin{table}[h!]
\caption{The time required for various stages shown in Figure~\ref{fig:overview} and the baseline experiments. KD refers to Knowledge Distillation, synchronous refers to fine-tuning using FedAvg, asynchronous refers to fine-tuning using asynchronous federated optimization. The fine-tuning is performed using ResNet-18, distilled from ResNet-34 via a ResNet-26 Teaching assistant.}
\label{tab:train}
% \vskip 0.15in
\begin{center}
\begin{small}
\begin{sc}
\resizebox{\columnwidth}{!}{
\begin{tabular}{lcccr}
\toprule
Dataset & Task & Epochs & Time\\
\midrule
Kinetics    & Train from scratch& 200& 31 hrs 26 mins\\
Kinetics & KD (No TA)& 200& 44 hrs 58 mins\\
Kinetics  & KD (TA)& 200 &  55 hrs 23 mins \\
HMDB51 & Fine-tune no clients & 80 & 3hrs 15mins\\
HMDB51    & Synchronous& 80 & 10 hrs 54 mins\\
HMDB51     & Asynchronous & 80& 6 hrs 31 mins\\
UCF101 & Fine-tune no clients & 80 & 22 hrs 5 mins\\
UCF101    & Synchronous& 80 & 74 hrs 27 mins\\
UCf101     & Asynchronous & 80& 44 hrs 7 mins\\
\bottomrule
\end{tabular}}
\end{sc}
\end{small}
\end{center}
\vskip -0.1in
\end{table}

As can be seen in Table~\ref{tab:clip_video}, the asynchronous training achieves higher accuracy for both per-clip and per-video metrics. This emphasizes the importance of our design of dealing with different rates of progress of different clients in the asynchronous training approach. 
From Table~\ref{tab:clip_video} we also see that 
our model performs better on UCF101 compared to HMDB51. HMDB51 contains several categories about different facial movements like smiling, laughing, chewing, and several other categories like eating and drinking. Such categories are not present in the list of categories of the data set UCF101 and are among the most difficult categories to deal with. Current state-of-the-art for UCF101 and HMDB51 stand at per-video top-1 accuracy of 98.69\% and 85.10\% respectively and are produced by the same work~\cite{kalfaoglu2020late}. Hence, it is evident that the HMDB51 is a difficult dataset to classify for deep learning architectures.
\begin{table}[h!]
\caption{The top-1 per-clip accuracy and the top-1 per-video accuracy}
\label{tab:clip_video}
% \vskip 0.15in
\begin{center}
\begin{small}
\begin{sc}
\resizebox{\columnwidth}{!}{
\begin{tabular}{lcccr}
\toprule
Dataset & Task & Per-Clip & Per-Video\\
\midrule
HMDB51 & Fine-tune no clients & 57.3\% & 64.1\%\\
HMDB51    & Synchronous& 54.4\% & 61.8\%\\
HMDB51     & Asynchronous & 55.6\% & 62.3\%\\
UCF101 & Fine-tune no clients & 85.7\% & 91.1\%\\
UCF101    & Synchronous& 84.3\% & 89.3\%\\
UCf101     & Asynchronous & 84.4\% & 89.5\%\\
\bottomrule
\end{tabular}}
\end{sc}
\end{small}
\end{center}
\vskip -0.1in
\end{table}

\subsection{Asynchronous Learning Hyperparamters}

\begin{figure}[h!]
% \vskip 0.2in
\begin{center}
\centerline{\includegraphics[width=0.9\columnwidth]{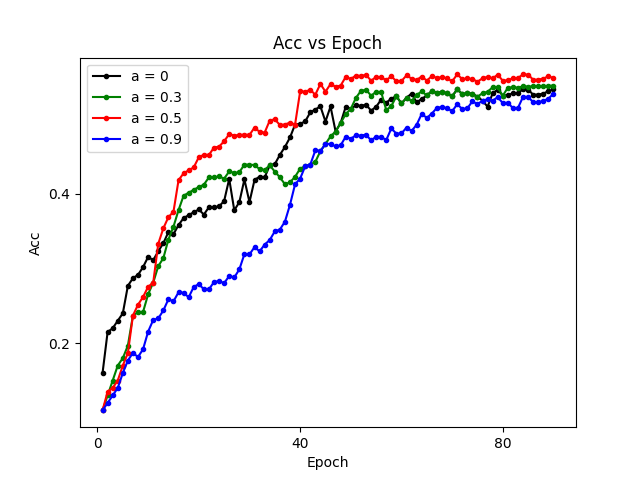}}
\caption{The function $s(t-\tau) = (1+t-\tau)^{-a}$ takes as an input the ``staleness" $t-\tau$. We set the mixing hyperparameter $\beta=0.7$ and vary the hyperparameter $a$. The asynchronous federated aggregation at the central server is adaptive; $\beta_t = \beta \times s(t-\tau)$ and $w_t = (1-\beta_{t})w_{t-1}+\beta_{t} w_{new}$. The figure depicts Asynchronous federated optimization performed on HMDB51.}
\label{fig:a_HMDB51}
\end{center}
\vskip -0.2in
\end{figure}

\begin{figure}[h!]
% \vskip 0.2in
\begin{center}
\centerline{\includegraphics[width=0.9\columnwidth]{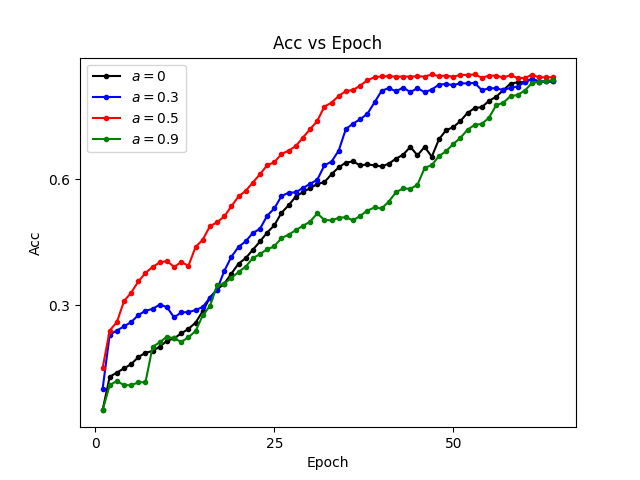}}
\caption{Asynchronous federated optimization performed on UCF101. We set the mixing hyperparameter $\beta=0.7$ and vary the hyperparameter $a$.}
\label{fig:a_UCF101}
\end{center}
\vskip -0.2in
\end{figure}

The asynchronous algorithm that we use adaptively updates the global model using the mixing hyperparameter $\beta$ and the function $s(t-\tau)=(1+t-\tau)^{-a}$. The function $s(\cdot)$ is monotonically decreasing with the staleness $t-\tau$. The intuition behind this is that in the time required for one client to perform local training and send its updates to the central server, several aggregations may have been performed at the central server. Hence, the global model has already learned more compared to the outdated updates that the client sends in. We perform experiments in order to find the best combination of the hyperparameters $a$ and $\beta$. In Figure~\ref{fig:a_HMDB51}, we keep $\beta = 0.7$ and vary $a$, here $a = 0$ refers to the case when we do not adjust the mixing parameter to account for stragglers: $\beta_{t} = \beta$. We see that $a=0.5$ is the best choice; not only is the convergence the fastest, the per-clip top-1 accuracy achieved for HMDB51 is the highest 55.6\% --- $a = 0$ gives an accuracy of 53.9\%, $a=0.3$ gives 54.2\% and $a=0.9$ gives 53.7\%. Therefore, we conclude that it's good to penalize clients for being late; however, keeping large penalties, such as $a=0.9$, will adversely affect speed of convergence since the weight assigned to updates received from clients will be very small during aggregation. Figure~\ref{fig:a_UCF101} shows the corresponding experiment for UCF101. We again observe that keeping $a=0.5$ gives the best results --- $a = 0$ and $a = 0.3$ give a per-clip accuracy of 83.7\%, $a = 0.5$ gives 84.4\% and $a = 0.9$ gives 83.6\%.
\begin{table}[h!]
\caption{ResNet-18 train times per epoch. For HMDB51 and UCF101, each client has approximately 500MB and 1.73GB of video data respectively.}
\label{tab:traintime}
% \vskip 0.15in
\begin{center}
\begin{small}
\begin{sc}
\resizebox{\columnwidth}{!}{
\begin{tabular}{lcccr}
\toprule
Dataset & Device & Train Time (per local epoch) \\
\midrule
HMDB51 & NVIDIA Jetson Nano & 391.1 seconds\\
HMDB51 & NVIDIA Jetson TX2 & 293.1 seconds \\
HMDB51 & NVIDIA Jetson Xavier NX & 121.3 seconds\\
HMDB51 & NVIDIA Jetson AGX Xavier & 84.5 seconds\\
UCF101 & NVIDIA Jetson Nano & 2691.6 seconds\\
UCF101 & NVIDIA Jetson TX2 & 2001.4 seconds \\
UCF101 & NVIDIA Jetson Xavier NX & 821.9 seconds\\
UCF101 & NVIDIA Jetson AGX Xavier & 572.1 seconds\\
\bottomrule
\end{tabular}
}
\end{sc}
\end{small}
\end{center}
\vskip -0.1in
\end{table}

\begin{figure}[h!]
% \vskip 0.2in
\begin{center}
\centerline{\includegraphics[width=0.9\columnwidth]{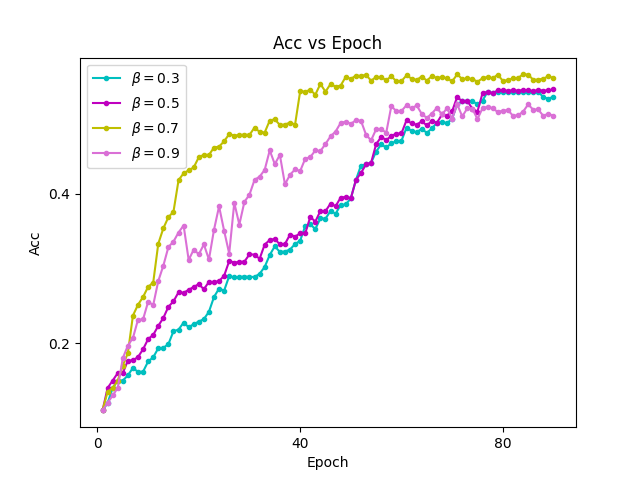}}
\caption{We set $a=0.5$ \& vary the mixing hyperparameter $\beta$. In the asynchronous federated optimization, convergence for small $\beta$ is slow as this corresponds to a smaller weight being assigned to the updated parameters received from the clients during aggregation. The figure depicts Asynchronous federated optimization performed on HMDB51.}
\label{fig:b_HMDB51}
\end{center}
\vskip -0.2in
\end{figure}

\begin{figure}[h!]
% \vskip 0.2in
\begin{center}
\centerline{\includegraphics[width=0.9\columnwidth]{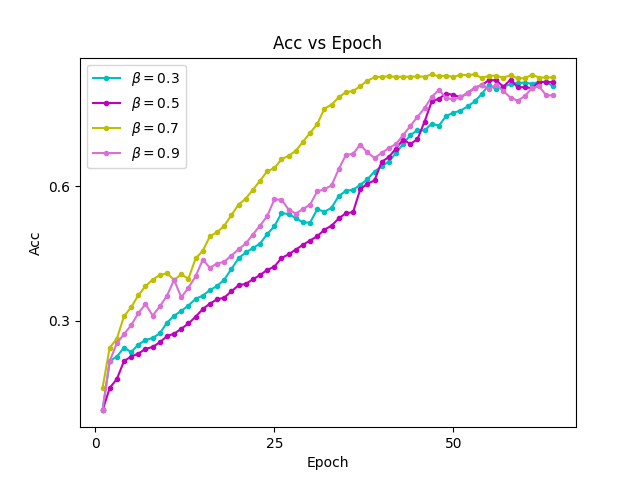}}
\caption{Asynchronous federated optimization performed on UCF101. We set $a=0.5$ \& vary the mixing hyperparameter $\beta$.}
\label{fig:b_UCF101}
\end{center}
\vskip -0.2in
\end{figure}

In the next set of experiments, we set $a = 0.5$ and vary $\beta$. For HMDB51, $\beta = 0.7$ gives the best per-clip accuracy of 55.6\%. Keeping a very low value assigns lower weight to the updates from the client during aggregation, and as can be seen in Figure~\ref{fig:b_HMDB51} the model's convergence slows down and leads to lower accuracy upon convergence --- $\beta = 0.3$ gives 53.6\%, $\beta = 0.5$ gives 53.8\%. A high value of the mixing parameter $\beta = 0.9$ gives 51.4\%. The low accuracy can be attributed to the fact that the dataset is distributed amongst many clients so assigning too high a weight to any client will lead to the model becoming biased. For the experiments on UCF101 as seen in Figure~\ref{fig:b_UCF101}, setting $a = 0.5$, $\beta = 0.7$ gives the best per-clip accuracy --- $\beta = 0.3$ gives 83.2\%, $\beta = 0.5$ gives 83.5\%, $\beta = 0.7$ gives 84.4\% and $\beta = 0.9$ gives 82.3\%. 

The combination of hyperparameters $a = 0.5$ and $\beta = 0.7$ gives the best results for both UCF101 and HMDB51. The {\bf top-1 per-video} setting $a = 0.5$ and $\beta = 0.7$ is 62.3\% and 89.5\% respectively. Table~\ref{tab:test} shows the inference times on the entire test dataset for the ResNet-18 architecture. From this and Table~\ref{tab:traintime}, it is clear that each client requires different amounts of time to train. For example, from Table~\ref{tab:traintime} we see that the training time per epoch is 4.7X more expensive on the Jetson Nano (the lowest spec of the four kinds of embedded devices) compared to the AGX Xavier (the highest spec). Therefore, this emphasizes the need for an asynchronous federated learning strategy rather than a synchronous one when dealing with heterogeneous embedded devices.

\begin{table}[h!]
\caption{ResNet-18 evaluated on the entire test dataset. The device heterogeneity is reflected in the inference times.}
\label{tab:test}
% \vskip 0.15in
\begin{center}
\begin{small}
\begin{sc}
\begin{tabular}{lcccr}
\toprule
Dataset & Device & Test Time \\
\midrule
HMDB51 & NVIDIA Jetson Nano & 181.4 seconds\\
HMDB51 & NVIDIA Jetson TX2 & 116.3 seconds \\
HMDB51 & NVIDIA Jetson Xavier NX & 89.4 seconds\\
HMDB51 & NVIDIA Jetson AGX Xavier & 68.3 seconds\\
UCF101 & NVIDIA Jetson Nano & 621.3 seconds\\
UCF101 & NVIDIA Jetson TX2 & 381.2 seconds \\
UCF101 & NVIDIA Jetson Xavier NX & 322.5 seconds\\
UCF101 & NVIDIA Jetson AGX Xavier & 217.7 seconds\\
\bottomrule
\end{tabular}
\end{sc}
\end{small}
\end{center}
\vskip -0.1in
\end{table}
\section{Discussion}
\label{sec:conclusion}
Our aim is to enable edge devices to perform action recognition, a computationally heavy task. Since in the real world, there is a mix of heterogeneous devices at the edge, we set out to develop our solution for such heterogeneity. Action recognition models have large datasets needed to learn complex spatio-temporal features---Kinetics-400 is approximately 400GB in size---and edge devices do not have this much disk space available. The largest from amongst our clients, NVIDIA Jetson AGX Xavier, has 32GB of storage. Datasets present on the edge devices, such as the HMDB51 (2GB), result in the ResNet-18 model overfitting (per-clip, top-1 accuracy of only 17.1\%). The solution to this is transfer learning: pre-training on the Kinetics dataset and fine-tuning on smaller datasets such as HMDB51 and UCF101. Our experiments on Kinetics show that a ResNet-18, trained from scratch, achieves a per-clip top-1 accuracy of 50.2\%, ResNet-18 distilled directly from ResNet-34 gives an accuracy of 53.8\%, and ResNet-18 distilled from a trained ResNet-34 via a teaching assistant (TA) achieves an accuracy of 54.6\%. Consequently, we use the ResNet-18 model distilled via a TA for fine-tuning. Furthermore, our experiments show that using the increment in accuracy from no TA to one TA is appreciable, however; adding more TAs results in negligible change in accuracy but increases the training time. 
Next, we experiment with federated optimization for a heterogeneous set of client devices. Heterogeneous devices pose a challenge for synchronous optimization: such aggregation will happen only after all devices send their updates. We therefore propose to use asynchronous federated averaging for our target scenario of heterogeneous embedded devices. We start off by analytically proving the convergence bound for our asynchronous approach. We empirically see that our asynchronous federated training takes 40\% less time than its synchronous counterpart. We also perform hyperparameter tuning of the asynchronous algorithm and determine that $a = 0.5$ and $\beta = 0.7$ gives the best accuracy for both HMBD51 and UCF101, which is not far behind the central server fine-tuning with no clients---for HMDB51, the per-clip accuracy for the federated asynchronous optimization is 55.6\%, as compared to 57.3\% on the central server with no clients. Thus, we for the first time empirically show that it is possible to achieve activity recognition on edge devices that are available available today.

In future work, one may consider how to handle non-iid data at the different clients. One should also consider dynamic ways of partitioning the data based on changing conditions like network bandwidth or availability or resource availability on each device.

\bibliography{main}

\begin{thebibliography}{10}
\providecommand{\url}[1]{#1}
\csname url@samestyle\endcsname
\providecommand{\newblock}{\relax}
\providecommand{\bibinfo}[2]{#2}
\providecommand{\BIBentrySTDinterwordspacing}{\spaceskip=0pt\relax}
\providecommand{\BIBentryALTinterwordstretchfactor}{4}
\providecommand{\BIBentryALTinterwordspacing}{\spaceskip=\fontdimen2\font plus
\BIBentryALTinterwordstretchfactor\fontdimen3\font minus
  \fontdimen4\font\relax}
\providecommand{\BIBforeignlanguage}[2]{{%
\expandafter\ifx\csname l@#1\endcsname\relax
\typeout{** WARNING: IEEEtran.bst: No hyphenation pattern has been}%
\typeout{** loaded for the language `#1'. Using the pattern for}%
\typeout{** the default language instead.}%
\else
\language=\csname l@#1\endcsname
\fi
#2}}
\providecommand{\BIBdecl}{\relax}
\BIBdecl

\bibitem{deng2009imagenet}
J.~Deng, W.~Dong, R.~Socher, L.-J. Li, K.~Li, and L.~Fei-Fei, ``Imagenet: A
  large-scale hierarchical image database,'' in \emph{2009 IEEE conference on
  computer vision and pattern recognition}.\hskip 1em plus 0.5em minus
  0.4em\relax Ieee, 2009, pp. 248--255.

\bibitem{yurur2014survey}
O.~Yurur, C.~H. Liu, and W.~Moreno, ``A survey of context-aware middleware
  designs for human activity recognition,'' \emph{IEEE Communications
  Magazine}, vol.~52, no.~6, pp. 24--31, 2014.

\bibitem{ranasinghe2016review}
S.~Ranasinghe, F.~Al~Machot, and H.~C. Mayr, ``A review on applications of
  activity recognition systems with regard to performance and evaluation,''
  \emph{International Journal of Distributed Sensor Networks}, vol.~12, no.~8,
  p. 1550147716665520, 2016.

\bibitem{girdhar2017actionvlad}
R.~Girdhar, D.~Ramanan, A.~Gupta, J.~Sivic, and B.~Russell, ``Actionvlad:
  Learning spatio-temporal aggregation for action classification,'' in
  \emph{Proceedings of the IEEE Conference on Computer Vision and Pattern
  Recognition}, 2017, pp. 971--980.

\bibitem{carreira2017quo}
J.~Carreira and A.~Zisserman, ``Quo vadis, action recognition? a new model and
  the kinetics dataset,'' in \emph{proceedings of the IEEE Conference on
  Computer Vision and Pattern Recognition}, 2017, pp. 6299--6308.

\bibitem{diba2018temporal}
A.~Diba, M.~Fayyaz, V.~Sharma, A.~Hossein~Karami, M.~Mahdi~Arzani,
  R.~Yousefzadeh, and L.~Van~Gool, ``Temporal 3d convnets using temporal
  transition layer,'' in \emph{Proceedings of the IEEE Conference on Computer
  Vision and Pattern Recognition Workshops}, 2018, pp. 1117--1121.

\bibitem{girdhar2019video}
R.~Girdhar, J.~Carreira, C.~Doersch, and A.~Zisserman, ``Video action
  transformer network,'' in \emph{Proceedings of the IEEE/CVF Conference on
  Computer Vision and Pattern Recognition}, 2019, pp. 244--253.

\bibitem{zhang2020few}
H.~Zhang, L.~Zhang, X.~Qi, H.~Li, P.~H. Torr, and P.~Koniusz, ``Few-shot action
  recognition with permutation-invariant attention,'' in \emph{Proceedings of
  the European Conference on Computer Vision (ECCV)}.\hskip 1em plus 0.5em
  minus 0.4em\relax Springer, 2020.

\bibitem{kumar2019protogan}
S.~Kumar~Dwivedi, V.~Gupta, R.~Mitra, S.~Ahmed, and A.~Jain, ``Protogan:
  Towards few shot learning for action recognition,'' in \emph{Proceedings of
  the IEEE/CVF International Conference on Computer Vision Workshops}, 2019,
  pp. 0--0.

\bibitem{brattoli2020rethinking}
B.~Brattoli, J.~Tighe, F.~Zhdanov, P.~Perona, and K.~Chalupka, ``Rethinking
  zero-shot video classification: End-to-end training for realistic
  applications,'' in \emph{Proceedings of the IEEE/CVF Conference on Computer
  Vision and Pattern Recognition}, 2020, pp. 4613--4623.

\bibitem{mandal2019out}
D.~Mandal, S.~Narayan, S.~K. Dwivedi, V.~Gupta, S.~Ahmed, F.~S. Khan, and
  L.~Shao, ``Out-of-distribution detection for generalized zero-shot action
  recognition,'' in \emph{Proceedings of the IEEE/CVF Conference on Computer
  Vision and Pattern Recognition}, 2019, pp. 9985--9993.

\bibitem{kay2017kinetics}
W.~Kay, J.~Carreira, K.~Simonyan, B.~Zhang, C.~Hillier, S.~Vijayanarasimhan,
  F.~Viola, T.~Green, T.~Back, P.~Natsev \emph{et~al.}, ``The kinetics human
  action video dataset,'' \emph{arXiv preprint arXiv:1705.06950}, 2017.

\bibitem{kuehne2011hmdb}
H.~Kuehne, H.~Jhuang, E.~Garrote, T.~Poggio, and T.~Serre, ``Hmdb: a large
  video database for human motion recognition,'' in \emph{2011 International
  conference on computer vision}.\hskip 1em plus 0.5em minus 0.4em\relax IEEE,
  2011, pp. 2556--2563.

\bibitem{soomro2012ucf101}
K.~Soomro, A.~R. Zamir, and M.~Shah, ``Ucf101: A dataset of 101 human actions
  classes from videos in the wild,'' \emph{arXiv preprint arXiv:1212.0402},
  2012.

\bibitem{hara2017learning}
K.~Hara, H.~Kataoka, and Y.~Satoh, ``Learning spatio-temporal features with 3d
  residual networks for action recognition,'' in \emph{Proceedings of the IEEE
  International Conference on Computer Vision Workshops}, 2017, pp. 3154--3160.

\bibitem{hara2018can}
------, ``Can spatiotemporal 3d cnns retrace the history of 2d cnns and
  imagenet?'' in \emph{Proceedings of the IEEE conference on Computer Vision
  and Pattern Recognition}, 2018, pp. 6546--6555.

\bibitem{li2019budgeted}
M.~Li, E.~Yumer, and D.~Ramanan, ``Budgeted training: Rethinking deep neural
  network training under resource constraints,'' \emph{arXiv preprint
  arXiv:1905.04753}, 2019.

\bibitem{bonawitz2019towards}
K.~Bonawitz, H.~Eichner, W.~Grieskamp, D.~Huba, A.~Ingerman, V.~Ivanov,
  C.~Kiddon, J.~Kone{\v{c}}n{\`y}, S.~Mazzocchi, H.~B. McMahan \emph{et~al.},
  ``Towards federated learning at scale: System design,'' \emph{arXiv preprint
  arXiv:1902.01046}, 2019.

\bibitem{feng2019computer}
X.~Feng, Y.~Jiang, X.~Yang, M.~Du, and X.~Li, ``Computer vision algorithms and
  hardware implementations: A survey,'' \emph{Integration}, vol.~69, pp.
  309--320, 2019.

\bibitem{bonawitz2017practical}
K.~Bonawitz, V.~Ivanov, B.~Kreuter, A.~Marcedone, H.~B. McMahan, S.~Patel,
  D.~Ramage, A.~Segal, and K.~Seth, ``Practical secure aggregation for
  privacy-preserving machine learning,'' in \emph{proceedings of the 2017 ACM
  SIGSAC Conference on Computer and Communications Security}, 2017, pp.
  1175--1191.

\bibitem{blanchard2017machine}
P.~Blanchard, E.~M. El~Mhamdi, R.~Guerraoui, and J.~Stainer, ``Machine learning
  with adversaries: Byzantine tolerant gradient descent,'' in \emph{Proceedings
  of the 31st International Conference on Neural Information Processing
  Systems}, 2017, pp. 118--128.

\bibitem{xu2019verifynet}
G.~Xu, H.~Li, S.~Liu, K.~Yang, and X.~Lin, ``Verifynet: Secure and verifiable
  federated learning,'' \emph{IEEE Transactions on Information Forensics and
  Security}, vol.~15, pp. 911--926, 2019.

\bibitem{mugunthan2020privacyfl}
V.~Mugunthan, A.~Peraire-Bueno, and L.~Kagal, ``Privacyfl: A simulator for
  privacy-preserving and secure federated learning,'' in \emph{Proceedings of
  the 29th ACM International Conference on Information \& Knowledge
  Management}, 2020, pp. 3085--3092.

\bibitem{chai2019towards}
Z.~Chai, H.~Fayyaz, Z.~Fayyaz, A.~Anwar, Y.~Zhou, N.~Baracaldo, H.~Ludwig, and
  Y.~Cheng, ``Towards taming the resource and data heterogeneity in federated
  learning,'' in \emph{2019 $\{$USENIX$\}$ Conference on Operational Machine
  Learning (OpML 19)}, 2019, pp. 19--21.

\bibitem{chai2020tifl}
Z.~Chai, A.~Ali, S.~Zawad, S.~Truex, A.~Anwar, N.~Baracaldo, Y.~Zhou,
  H.~Ludwig, F.~Yan, and Y.~Cheng, ``Tifl: A tier-based federated learning
  system,'' in \emph{Proceedings of the 29th International Symposium on
  High-Performance Parallel and Distributed Computing}, 2020, pp. 125--136.

\bibitem{xie2019asynchronous}
C.~Xie, S.~Koyejo, and I.~Gupta, ``Asynchronous federated optimization,''
  \emph{arXiv preprint arXiv:1903.03934}, 2019.

\bibitem{hinton2015distilling}
G.~Hinton, O.~Vinyals, and J.~Dean, ``Distilling the knowledge in a neural
  network,'' \emph{NIPS 2014 Deep Learning Workshop}, 2015.

\bibitem{mirzadeh2020improved}
S.~I. Mirzadeh, M.~Farajtabar, A.~Li, N.~Levine, A.~Matsukawa, and
  H.~Ghasemzadeh, ``Improved knowledge distillation via teacher assistant,'' in
  \emph{Proceedings of the AAAI Conference on Artificial Intelligence},
  vol.~34, no.~04, 2020, pp. 5191--5198.

\bibitem{liu2020fedvision}
Y.~Liu, A.~Huang, Y.~Luo, H.~Huang, Y.~Liu, Y.~Chen, L.~Feng, T.~Chen, H.~Yu,
  and Q.~Yang, ``Fedvision: An online visual object detection platform powered
  by federated learning,'' in \emph{Proceedings of the AAAI Conference on
  Artificial Intelligence}, vol.~34, no.~08, 2020, pp. 13\,172--13\,179.

\bibitem{mcmahan2017communication}
B.~McMahan, E.~Moore, D.~Ramage, S.~Hampson, and B.~A. y~Arcas,
  ``Communication-efficient learning of deep networks from decentralized
  data,'' in \emph{Artificial Intelligence and Statistics}.\hskip 1em plus
  0.5em minus 0.4em\relax PMLR, 2017, pp. 1273--1282.

\bibitem{yu2019federated}
P.~Yu and Y.~Liu, ``Federated object detection: Optimizing object detection
  model with federated learning,'' in \emph{Proceedings of the 3rd
  International Conference on Vision, Image and Signal Processing}, 2019, pp.
  1--6.

\bibitem{kaissis2020secure}
G.~A. Kaissis, M.~R. Makowski, D.~R{\"u}ckert, and R.~F. Braren, ``Secure,
  privacy-preserving and federated machine learning in medical imaging,''
  \emph{Nature Machine Intelligence}, vol.~2, no.~6, pp. 305--311, 2020.

\bibitem{jeong2018communication}
E.~Jeong, S.~Oh, H.~Kim, J.~Park, M.~Bennis, and S.-L. Kim,
  ``Communication-efficient on-device machine learning: Federated distillation
  and augmentation under non-iid private data,'' \emph{arXiv preprint
  arXiv:1811.11479}, 2018.

\bibitem{anil2018large}
R.~Anil, G.~Pereyra, A.~Passos, R.~Ormandi, G.~E. Dahl, and G.~E. Hinton,
  ``Large scale distributed neural network training through online
  distillation,'' \emph{arXiv preprint arXiv:1804.03235}, 2018.

\bibitem{lecun1998gradient}
Y.~LeCun, L.~Bottou, Y.~Bengio, and P.~Haffner, ``Gradient-based learning
  applied to document recognition,'' \emph{Proceedings of the IEEE}, vol.~86,
  no.~11, pp. 2278--2324, 1998.

\bibitem{sozinov2018human}
K.~Sozinov, V.~Vlassov, and S.~Girdzijauskas, ``Human activity recognition
  using federated learning,'' in \emph{2018 IEEE Intl Conf on Parallel \&
  Distributed Processing with Applications, Ubiquitous Computing \&
  Communications, Big Data \& Cloud Computing, Social Computing \& Networking,
  Sustainable Computing \& Communications
  (ISPA/IUCC/BDCloud/SocialCom/SustainCom)}.\hskip 1em plus 0.5em minus
  0.4em\relax IEEE, 2018, pp. 1103--1111.

\bibitem{ek2020evaluation}
S.~Ek, F.~Portet, P.~Lalanda, and G.~Vega, ``Evaluation of federated learning
  aggregation algorithms: application to human activity recognition,'' in
  \emph{Adjunct Proceedings of the 2020 ACM International Joint Conference on
  Pervasive and Ubiquitous Computing and Proceedings of the 2020 ACM
  International Symposium on Wearable Computers}, 2020, pp. 638--643.

\bibitem{wang2019adaptive}
S.~Wang, T.~Tuor, T.~Salonidis, K.~K. Leung, C.~Makaya, T.~He, and K.~Chan,
  ``Adaptive federated learning in resource constrained edge computing
  systems,'' \emph{IEEE Journal on Selected Areas in Communications}, vol.~37,
  no.~6, pp. 1205--1221, 2019.

\bibitem{lim2020federated}
W.~Y.~B. Lim, N.~C. Luong, D.~T. Hoang, Y.~Jiao, Y.-C. Liang, Q.~Yang,
  D.~Niyato, and C.~Miao, ``Federated learning in mobile edge networks: A
  comprehensive survey,'' \emph{IEEE Communications Surveys \& Tutorials},
  vol.~22, no.~3, pp. 2031--2063, 2020.

\bibitem{xu2019elfish}
Z.~Xu, Z.~Yang, J.~Xiong, J.~Yang, and X.~Chen, ``Elfish: Resource-aware
  federated learning on heterogeneous edge devices,'' \emph{arXiv preprint
  arXiv:1912.01684}, 2019.

\bibitem{chen2019asynchronous}
Y.~Chen, Y.~Ning, M.~Slawski, and H.~Rangwala, ``Asynchronous online federated
  learning for edge devices with non-iid data,'' \emph{arXiv preprint
  arXiv:1911.02134}, 2019.

\bibitem{nikouei2018smart}
S.~Y. Nikouei, Y.~Chen, S.~Song, R.~Xu, B.-Y. Choi, and T.~Faughnan, ``Smart
  surveillance as an edge network service: From harr-cascade, svm to a
  lightweight cnn,'' in \emph{2018 IEEE 4th International Conference on
  Collaboration and Internet Computing (CIC)}.\hskip 1em plus 0.5em minus
  0.4em\relax IEEE, 2018, pp. 256--265.

\bibitem{lin2018edgespeechnets}
Z.~Q. Lin, A.~G. Chung, and A.~Wong, ``Edgespeechnets: Highly efficient deep
  neural networks for speech recognition on the edge,'' \emph{arXiv preprint
  arXiv:1810.08559}, 2018.

\bibitem{reisizadeh2020fedpaq}
A.~Reisizadeh, A.~Mokhtari, H.~Hassani, A.~Jadbabaie, and R.~Pedarsani,
  ``Fedpaq: A communication-efficient federated learning method with periodic
  averaging and quantization,'' in \emph{International Conference on Artificial
  Intelligence and Statistics}.\hskip 1em plus 0.5em minus 0.4em\relax PMLR,
  2020, pp. 2021--2031.

\bibitem{sattler2019robust}
F.~Sattler, S.~Wiedemann, K.-R. M{\"u}ller, and W.~Samek, ``Robust and
  communication-efficient federated learning from non-iid data,'' \emph{IEEE
  transactions on neural networks and learning systems}, vol.~31, no.~9, pp.
  3400--3413, 2019.

\bibitem{mills2019communication}
J.~Mills, J.~Hu, and G.~Min, ``Communication-efficient federated learning for
  wireless edge intelligence in iot,'' \emph{IEEE Internet of Things Journal},
  vol.~7, no.~7, pp. 5986--5994, 2019.

\bibitem{smith2017federated}
V.~Smith, C.-K. Chiang, M.~Sanjabi, and A.~Talwalkar, ``Federated multi-task
  learning,'' in \emph{Proceedings of the 31st International Conference on
  Neural Information Processing Systems (NeurIPS)}, 2017, pp. 4427--4437.

\bibitem{zhang2018shufflenet}
X.~Zhang, X.~Zhou, M.~Lin, and J.~Sun, ``Shufflenet: An extremely efficient
  convolutional neural network for mobile devices,'' in \emph{Proceedings of
  the IEEE conference on computer vision and pattern recognition}, 2018, pp.
  6848--6856.

\bibitem{wu2019fbnet}
B.~Wu, X.~Dai, P.~Zhang, Y.~Wang, F.~Sun, Y.~Wu, Y.~Tian, P.~Vajda, Y.~Jia, and
  K.~Keutzer, ``Fbnet: Hardware-aware efficient convnet design via
  differentiable neural architecture search,'' in \emph{Proceedings of the
  IEEE/CVF Conference on Computer Vision and Pattern Recognition}, 2019, pp.
  10\,734--10\,742.

\bibitem{han2020ghostnet}
K.~Han, Y.~Wang, Q.~Tian, J.~Guo, C.~Xu, and C.~Xu, ``Ghostnet: More features
  from cheap operations,'' in \emph{Proceedings of the IEEE/CVF Conference on
  Computer Vision and Pattern Recognition}, 2020, pp. 1580--1589.

\bibitem{jang2020knowledge}
I.~Jang, H.~Kim, D.~Lee, Y.-S. Son, and S.~Kim, ``Knowledge transfer for
  on-device deep reinforcement learning in resource constrained edge computing
  systems,'' \emph{IEEE Access}, vol.~8, pp. 146\,588--146\,597, 2020.

\bibitem{matsubara2020head}
Y.~Matsubara, D.~Callegaro, S.~Baidya, M.~Levorato, and S.~Singh, ``Head
  network distillation: Splitting distilled deep neural networks for
  resource-constrained edge computing systems,'' \emph{IEEE Access}, vol.~8,
  pp. 212\,177--212\,193, 2020.

\bibitem{he2019knowledge}
T.~He, C.~Shen, Z.~Tian, D.~Gong, C.~Sun, and Y.~Yan, ``Knowledge adaptation
  for efficient semantic segmentation,'' in \emph{Proceedings of the IEEE/CVF
  Conference on Computer Vision and Pattern Recognition}, 2019, pp. 578--587.

\bibitem{liu2019semantic}
Q.~Liu, L.~Xie, H.~Wang, and A.~L. Yuille, ``Semantic-aware knowledge
  preservation for zero-shot sketch-based image retrieval,'' in
  \emph{Proceedings of the IEEE/CVF International Conference on Computer
  Vision}, 2019, pp. 3662--3671.

\bibitem{xie2020self}
Q.~Xie, M.-T. Luong, E.~Hovy, and Q.~V. Le, ``Self-training with noisy student
  improves imagenet classification,'' in \emph{Proceedings of the IEEE/CVF
  Conference on Computer Vision and Pattern Recognition}, 2020, pp.
  10\,687--10\,698.

\bibitem{thoker2019cross}
F.~M. Thoker and J.~Gall, ``Cross-modal knowledge distillation for action
  recognition,'' in \emph{2019 IEEE International Conference on Image
  Processing (ICIP)}.\hskip 1em plus 0.5em minus 0.4em\relax IEEE, 2019, pp.
  6--10.

\bibitem{he2016deep}
K.~He, X.~Zhang, S.~Ren, and J.~Sun, ``Deep residual learning for image
  recognition,'' in \emph{Proceedings of the IEEE conference on computer vision
  and pattern recognition}, 2016, pp. 770--778.

\bibitem{wang2018deep}
M.~Wang and W.~Deng, ``Deep visual domain adaptation: A survey,''
  \emph{Neurocomputing}, vol. 312, pp. 135--153, 2018.

\bibitem{sun2019patient}
S.~Sun, Y.~Cheng, Z.~Gan, and J.~Liu, ``Patient knowledge distillation for bert
  model compression,'' in \emph{Proceedings of the 2019 Conference on Empirical
  Methods in Natural Language Processing and the 9th International Joint
  Conference on Natural Language Processing (EMNLP-IJCNLP)}, 2019, pp.
  4314--4323.

\bibitem{liu2017sparse}
J.~Liu, Y.~Wang, and Y.~Qiao, ``Sparse deep transfer learning for convolutional
  neural network,'' in \emph{Proceedings of the AAAI Conference on Artificial
  Intelligence}, vol.~31, no.~1, 2017.

\bibitem{kalfaoglu2020late}
M.~E. Kalfaoglu, S.~Kalkan, and A.~A. Alatan, ``Late temporal modeling in 3d
  cnn architectures with bert for action recognition,'' in \emph{European
  Conference on Computer Vision}.\hskip 1em plus 0.5em minus 0.4em\relax
  Springer, 2020, pp. 731--747.

\end{thebibliography}
\bibliographystyle{icml2021}

\end{document}